\newcommand*{\ATLASLATEXPATH}{latex/}
\documentclass[UKenglish,texlive=2011,cernpreprint,txfonts]{\ATLASLATEXPATH atlasdoc}

\usepackage[biblatex=false,siunitx=false]{\ATLASLATEXPATH atlaspackage}
\usepackage{\ATLASLATEXPATH atlasbiblatex}

\usepackage{\ATLASLATEXPATH atlascontribute}

\usepackage{\ATLASLATEXPATH atlasphysics}

\graphicspath{{logos/}{figures/}}

\usepackage{cite}

\usepackage{sbottom_Run2-defs}

\AtlasTitle{Search for bottom squark pair production in proton--proton collisions at $\sqrt{s}=13$~\TeV\ with the ATLAS detector}

\author{The ATLAS Collaboration}

\AtlasRefCode{SUSY-2015-01}

\PreprintIdNumber{CERN-PH-2016-138}

\AtlasJournalRef{Eur. Phys. J. C (2016) 76:547}
\AtlasDOI{10.1140/epjc/s10052-016-4382-4}

\AtlasAbstract{%
The result of a search for pair production of the supersymmetric partner of the Standard Model bottom quark ($\bone$) is reported. 
The search uses \intlumi~of $pp$ collisions at $\sqrt{s}=13$~\TeV\ collected by the ATLAS experiment at the Large Hadron Collider in 2015. 
Bottom squarks are searched for in events containing large missing transverse momentum and exactly two jets identified as 
originating from $b$-quarks. No excess above the expected Standard Model background yield is observed. Exclusion limits at 95\% confidence 
level on the mass of the bottom squark are derived in phenomenological supersymmetric $R$-parity-conserving models in which the $\bone$ is the 
lightest squark and is assumed to decay exclusively via $\bone \rightarrow b \neut$, where \neut~is the lightest neutralino. 
The limits significantly extend previous results; bottom squark masses up to 800 (840)~\GeV\ are excluded for the $\neut$ mass below 360 (100)~\GeV\ whilst differences in mass 
above 100~\GeV\ between the \bone and the \neut~are excluded up to a $\bone$ mass of 500~\GeV. }

\hypersetup{pdftitle={ATLAS document},pdfauthor={The ATLAS Collaboration}}

\begin{document}

\maketitle


%
%
\newcommand{\NumUnc}[2]{\makebox[4ex]{\hfill #1} \pm \makebox[4ex]{#2  \hfill}}
\section{Introduction}
\label{sec:intro}
Supersymmetry (SUSY)~\cite{Golfand:1971iw,Volkov:1973ix,Wess:1974tw,Wess:1974jb,Ferrara:1974pu,Salam:1974ig} provides an extension of the Standard Model (SM) that solves the hierarchy problem~\cite{Sakai:1981gr,Dimopoulos:1981yj,Ibanez:1981yh,Dimopoulos:1981zb} by introducing supersymmetric partners of  the known bosons and fermions. In the framework of the $R$-parity-conserving minimal supersymmetric extension of the SM (MSSM)~\cite{Fayet:1976et,Fayet:1977yc,Farrar:1978xj}, SUSY particles are produced in pairs and the lightest supersymmetric particle (LSP) is stable, providing a possible candidate for dark matter~\cite{Goldberg:1983nd,Ellis:1983ew}. In a large variety of models the LSP is the lightest neutralino (\neut). 
Naturalness considerations~\cite{Barbieri:1987fn,deCarlos:1993yy} suggest that the supersymmetric partners of the third-generation SM quarks are the lightest coloured supersymmetric particles. This may lead to the lightest bottom squark ($\bone$) and top squark ($\tone$) mass eigenstates\footnote{Scalar partners of the left-handed and right-handed chiral components of the bottom quark ($\tilde{b}_{\mathrm{L},\mathrm{R}}$) or top quark ($\tilde{t}_{\mathrm{L},\mathrm{R}}$) mix to form mass eigenstates for which \bone and \tone are defined as the lighter of the two.} being significantly lighter than the other squarks and the gluinos. As a consequence, $\bone$ and $\tone$ could be pair-produced with relatively large cross-sections at the Large Hadron Collider (LHC). 

This paper presents a search for the pair production of bottom squarks decaying exclusively
as $\bone \rightarrow b\neut$ using \intlumi~of proton--proton ($pp$) collisions at $\sqrt{s}=13$~\TeV\ collected by the ATLAS experiment at the LHC in 2015. A SUSY particle mass hierarchy with this exclusive decay has been predicted by various phenomenological MSSM models~\cite{SUSY-2014-08}.
Searches with the $\sqrt{s}=8$~\TeV\ LHC Run-1 dataset at ATLAS and CMS have set limits on $\bone$ masses in such scenarios. For $\neut$ masses around 100~\GeV, exclusion limits at 95\% confidence level (CL) up to 620~\GeV\ and 680~\GeV\ have been reported by the ATLAS~\cite{SUSY-2014-07} and CMS~\cite{CMS-SUS-14-001} collaborations, respectively. The searches were performed in events characterized by the presence of large missing transverse momentum 
and two jets identified as containing $b$-hadrons (\bjets). 
In Run 2 of the LHC the production cross-section rises due to the increase of the centre-of-mass energy of the $pp$ collisions. For instance, for a $\bone$ with a mass of 800~\GeV, the production cross-section increases by almost a factor of ten  going from $\sqrt{s}=8$~\TeV\ to $\sqrt{s}=13$~\TeV. In addition, the sensitivity of the analysis benefits from the improved algorithms adopted to identify $b$-jets 
and use of information from the newly installed pixel layer in the Run-2 ATLAS detector.   
The search strategy at 13~\TeV\ closely follows  the previous ATLAS studies, with signal regions defined to provide sensitivity to the kinematic topologies arising from different mass splittings  between the bottom squark and the neutralino.


\section{ATLAS detector}
\label{sec:detector}
The ATLAS detector~\cite{PERF-2007-01} is a multi-purpose particle physics detector with a forward-backward symmetric cylindrical
geometry and nearly 4$\pi$ coverage in solid angle.\footnote{ATLAS uses a right-handed coordinate system with its origin at the nominal
interaction point in the centre of the detector. The positive $x$-axis is defined by the direction from the interaction point to the centre
of the LHC ring, with the positive $y$-axis pointing upwards, while the beam direction defines the $z$-axis. Cylindrical coordinates 
$(r,\phi)$ are used in the transverse plane, $\phi$ being the azimuthal angle around the $z$-axis. The pseudorapidity $\eta$ is
defined in terms of the polar angle $\theta$ by $\eta=-\ln\tan(\theta/2)$. Rapidity is defined as $y$ = $0.5 \ln[(E + p_z)/(E-p_z)]$ where $E$ denotes the energy and $p_z$ is the component of the momentum along the beam direction.}
The inner tracking detector consists of pixel and silicon microstrip detectors 
covering the pseudorapidity region $|\eta|<2.5$, surrounded by a transition radiation tracker  
which enhances electron identification in the region $|\eta|<2.0$. 
Between Run 1 and Run 2, a new inner pixel layer, the Insertable B-Layer (IBL)\cite{IBL}, was inserted at a mean sensor radius of 
3.3~cm. The inner detector is surrounded by a thin superconducting solenoid providing an axial 2~T magnetic field and by a fine-granularity lead/liquid-argon (LAr) electromagnetic calorimeter covering $|\eta|<3.2$. A steel/scintillator-tile calorimeter provides hadronic coverage in the central pseudorapidity range ($|\eta|<1.7$). 
The endcap and forward regions ($1.5<|\eta|<4.9$) of the hadronic calorimeter are made of LAr active layers with either copper or tungsten 
as the absorber material. 
An extensive muon spectrometer with an air-core toroid magnet system surrounds the calorimeters.
Three layers of high-precision tracking chambers provide coverage in the range $|\eta|<2.7$, while dedicated fast chambers allow triggering in the region $|\eta|<2.4$. The ATLAS trigger system consists of a hardware-based level-1 trigger followed by a software-based high-level trigger~\cite{PERF-2011-02}.


\section{Data and simulated event samples}
\label{sec:MC}
The data used in this analysis were collected by the ATLAS detector in $pp$ collisions at the LHC with a centre-of-mass energy of 13~\TeV\ and 25 ns proton bunch crossing interval during 2015. After applying beam-, data- and detector-quality criteria, the available dataset corresponds to an integrated luminosity of \intlumi with an uncertainty of $\pm$5\%. The uncertainty is derived from a calibration of the luminosity scale using a pair of $x$--$y$ beam-separation scans performed in August 2015 and following a methodology similar to that detailed in \Refe~\cite{DAPR-2011-01}. In this dataset, each event includes an average of approximately 14 additional inelastic $pp$ collisions in the same bunch crossing (in-time pile-up). 
The events used in this search were selected using a trigger logic that accepts events with an uncorrected missing transverse momentum above 70~\GeV, calculated using a sum over calorimeter cells. Additional events selected with lepton- and photon-based triggers are employed for control regions defined to estimate SM background contributions. 
All the selections employed in this paper use a highly efficient trigger selection. The triggers used are in a plateau 
region and the systematic uncertainties related to the trigger simulation are found to be negligible.

Monte Carlo (MC) simulated event samples are used to model the expected signal and to aid in the description and estimation of SM background processes.  The response of the detector is simulated~\cite{SOFT-2010-01} either fully by a software program based on \geant~\cite{Agostinelli:2002hh} or by a faster simulation based on a parameterization~\cite{ATL-PHYS-PUB-2010-013} for the calorimeter response and \geant for the other detector systems. To account for additional $pp$ interactions from the same and close-by bunch crossings, a set of minimum-bias interactions generated using \pythia~\cite{Sjostrand:2007gs} 8.186 and the MSTW2008LO~\cite{Martin:2009iq,Martin:2007bv} parton distribution function (PDF) set is superimposed onto the hard scattering events to reproduce the observed distribution of the average number of interactions per bunch crossing.

The signal samples are generated using \madgraph~\cite{Alwall:2014hca} v2.2.3 interfaced to \pythia~8.186 with the A14~\cite{ATL-PHYS-PUB-2014-021} set of parameters (tune) for the modelling of the parton showering (PS), hadronization and underlying event. The matrix element (ME) calculation is performed at tree level and includes the emission of up to two additional partons. The PDF set used for the generation is NNPDF23LO~\cite{Ball:2012cx}. The ME--PS matching is done using the CKKW-L~\cite{Lonnblad:2012ix} prescription, with a matching scale set to one quarter of the bottom squark mass. The cross-sections used to evaluate the signal yields are calculated to next-to-leading-order accuracy in the strong coupling constant, adding the resummation of soft gluon emission at next-to-leading-logarithmic accuracy (NLO+NLL)~\cite{Beenakker:1997ut,Beenakker:2010nq,Beenakker:2011fu}. 

SM background samples are simulated using different MC generator programs depending on the process. 
Events containing $W$ or $Z$ bosons with associated jets, including jets from the fragmentation of 
$b$- and $c$-quarks (heavy-flavour jets), are simulated using 
the \sherpa 2.1.1~\cite{Gleisberg:2008ta} generator. Matrix elements are calculated 
for up to two additional partons at next-to-leading order (NLO) and four partons at leading order (LO) using the \comix~\cite{Gleisberg:2008fv} and 
\openloops~\cite{Cascioli:2011va} matrix element generators and merged with the \sherpa 
PS~\cite{Schumann:2007mg} using the ME+PS@NLO prescription~\cite{Hoeche:2012yf}. 
The CT10~\cite{Lai:2010vv} PDF set is used in conjunction with a dedicated PS tune developed 
by the \sherpa authors. The $W$/$Z$+jets events are normalized to their next-to-NLO (NNLO) QCD theoretical cross-sections~\cite{Gavin:2010az}.  For a cross-check of the modelling of \Zjets\ background, samples of events containing a photon produced in  association with jets, including heavy-flavour jets, are simulated using the \sherpa 2.1.1 generator with matrix elements calculated at LO with up to four partons for the LO calculation.

Diboson processes are also simulated using the \sherpa 2.1.1 generator with the same settings as the single-boson samples. 
They are calculated for up to one ($ZZ$) or zero ($WW, WZ$) additional partons at NLO and up to three additional partons at LO. 
The NLO generator cross-sections are scaled down by 9\% to account for the usage of $\alpha^{}_{\rm QED}$=1/129 rather 
than 1/132, the latter corresponding to the use of the parameters defined by the Particle Data Group as 
input to the $G^{}_{\mu}$ scheme~\cite{PDG:2014}.

Top-quark pair and single-top-quark production ($Wt$- and $s$-channel) events are simulated using the \powhegbox v2~\cite{Alioli:2010xd} 
generator as described in Ref.~\cite{ATL-PHYS-PUB-2016-004}
with the CT10 PDF set in the matrix element calculations. Electroweak $t$-channel single-top-quark events are generated using the \powhegbox v1 generator. 
The top-quark mass is set to 172.5~\GeV. For the \ttbar production, the $h^{}_{\rm damp}$ parameter, which controls the transverse momentum (\pt)~of the first additional emission beyond the Born configuration and thus regulates the \pt~of the recoil emission 
against the \ttbar~system, is set to the mass of the top quark. 
For all processes, the parton shower, fragmentation, and the underlying event are simulated using \pythia~6.428~\cite{Sjostrand:2006za} with the 
CTEQ6L1 PDF set and the corresponding Perugia 2012 tune (P2012)~\cite{Skands:2010ak}. 
The \ttbar\ samples are normalized to their NNLO cross-section including 
the resummation of soft gluon emission at next-to-NLL accuracy using \verb|Top++2.0|~\cite{Czakon:2011xx}.
Samples of single-top-quark production are normalized to the NLO cross-sections reported in \Refs~\cite{Kidonakis:2011wy,Kidonakis:2010tc,Kidonakis:2010ux} for the $s$-, $t$- and $Wt$-channels, respectively.
The associated production of a top-quark pair with a vector boson, $\ttV$, is generated at LO with \madgraph~v2.2.3 interfaced to \pythia~8.186, 
with up to two ($\ttW$), one ($\ttZ$) or no ($\ttWW$) extra partons included in the matrix elements. The samples are normalized to their NLO 
cross-sections~\cite{Alwall:2014hca}. 

The EvtGen 1.2.0 program~\cite{EvtGen} is used for modelling the properties of  bottom- and charm-hadron decays 
in all samples generated with \madgraph and \powhegbox.  

Several samples produced without detector simulation are employed to derive systematic uncertainties associated with the specific 
configuration of the MC generators used for the nominal SM background samples. They include variations of the renormalization and factorization scales, the CKKW-L 
matching scale, as well as different PDF sets and fragmentation/hadronization models. Details of the MC modelling uncertainties 
are discussed in \Sect~\ref{sec:systematic}.


\section{Event reconstruction}
\label{sec:objects}
The search for bottom squark pair production is based on a selection of events with jets and large missing transverse momentum in the final state. Events containing electrons or muons are explicitly vetoed in the signal and validation regions, and are used to define control regions. An overlap removal procedure is applied to prevent double-counting of reconstructed objects. The details of the selections and overlap removal are given below.

Interaction vertices from the $pp$ collisions are reconstructed from tracks in the inner detector. Events must have at least one primary reconstructed vertex, required to be consistent with the beamspot envelope and to consist of at least two tracks with \pt$>$~0.4~\GeV. When more than one such vertex is found the one with the largest  sum of the square of  transverse momenta of associated tracks~\cite{ATL-PHYS-PUB-2015-026} is chosen. 

Jet candidates are reconstructed from three-dimensional energy clusters~\cite{PERF-2014-07} in the calorimeter using the anti-$k_t$ jet algorithm~\cite{Cacciari:2008gp} with a radius parameter of 0.4. Each topological cluster's energy is calibrated to the electromagnetic scale prior to jet reconstruction. The reconstructed jets are then calibrated to the particle level by applying a jet energy scale (JES) derived from simulation and {\em in situ} corrections based on 8~\TeV\ data~\cite{ATL-PHYS-PUB-2015-015,ATLAS-CONF-2015-037}. Quality criteria are imposed to identify jets arising from non-collision sources or detector noise and any event containing such a jet is removed~\cite{ATLAS-CONF-2015-029}. Further track-based selections are applied to reject jets with $\pt<60$~\GeV\ and $|\eta|<2.4$ that originate from pile-up interactions~\cite{ATLAS-CONF-2014-018} and the expected event average energy contribution from pile-up clusters is subtracted using a factor dependent on the jet area~\cite{ATL-PHYS-PUB-2015-015}. Jets are classified as ``baseline'' and ``signal''. Baseline jets are required to have $\pt>20$~\GeV\ and $|\eta|<2.8$. Signal jets, selected after resolving overlaps with electrons and muons, are required to pass the stricter requirement of $\pt>35$~\GeV. 

Jets are identified as \bjets if tagged by a multivariate algorithm which uses information about the impact parameters of inner detector tracks matched to the jet, the presence of displaced secondary vertices, and the reconstructed flight paths of $b$- and $c$-hadrons inside the jet~\cite{ATL-PHYS-PUB-2015-039}. 
The \btagging working point with a 77\% efficiency, as determined in a simulated sample of \ttbar events, was chosen as part of the optimization process discussed in \Sect~\ref{sec:selection}. The corresponding rejection factors against jets originating from \cquarks and from light quarks and gluons at this working point are 4.5 and 130, respectively~\cite{ATLAS-CONF-2014-046}. To compensate for differences between data and MC simulation in the \btagging efficiencies and mis-tag rates, correction factors derived from data-driven methods are applied to the simulated samples~\cite{PERF-2012-04}.  Candidate $b$-tagged jets are required to have $\pt > 50$~\GeV\ and $|\eta| <$ 2.5.

Electron candidates are reconstructed from energy clusters in the electromagnetic calorimeter matched to a track in the  inner detector and are required to satisfy a set of ``loose'' quality criteria~\cite{PERF-2013-03,PERF-2013-05,ATL-PHYS-PUB-2015-041}. They are also required to lie within the fiducial volume $|\eta|<2.47$. Muon candidates are reconstructed from matching tracks in the inner detector and the muon spectrometer. Events containing one or more muon candidates that have a transverse (longitudinal) impact parameter with respect to the primary vertex larger than 0.2~mm (1~mm) are rejected to suppress cosmic rays. Muon candidates  are also required to satisfy ``medium'' quality criteria~\cite{PERF-2015-10} and have $|\eta |$ < 2.5. All electron and muon candidates must have $\pt>$ 10~\GeV. Lepton candidates remaining after resolving the overlap with baseline jets (see next paragraph) are called ``baseline'' leptons. In the control regions where lepton identification is required, ``signal'' leptons are chosen from the baseline set with \pt$>$~26~\GeV\ to ensure full efficiency of the trigger and are required to be isolated from other activity using a selection designed to accept 99\% of leptons from $Z$ boson decays.
Signal electrons are further required to satisfy ``tight'' quality criteria~\cite{PERF-2013-03}. Electrons (muons) are matched to the primary vertex by requiring the transverse impact parameter ($d_0$) to satisfy $|d_0/\sigma(d_0)|<$~5 (3), and the longitudinal impact parameter ($z_0$) to satisfy $|z_0 \sin \theta|<$~0.5~mm for both the electrons and muons. The MC events are corrected to account for differences in the
lepton trigger, reconstruction and identification efficiencies between
data and MC simulation.

The sequence to resolve overlapping objects begins by removing electron candidates sharing an inner detector track with a muon candidate. Next, jet candidates within $\Delta R=\sqrt{(\Delta y)^2 + (\Delta \phi)^2} =0.2$ of an electron candidate are discarded, unless the jet is \btagged, in which case the electron is discarded since it is likely to originate from a semileptonic \bhadron decay. Finally, any lepton candidate remaining within $\Delta R = 0.4$ of any surviving jet candidate is discarded, except for the case where the lepton is a muon and the number of tracks associated with the jet is less than three.

The missing transverse momentum \ptmissvec, with magnitude \etmiss, is defined as the negative vector sum of the \pt of all selected and calibrated physics objects in the event, with an extra term added to account for soft energy in the event which is not associated with any of the selected objects. This soft term is calculated from inner detector tracks with \pt above 400~\MeV\ matched to the primary vertex to make it more robust against pile-up contamination~\cite{ATLAS-CONF-2013-082,PERF-2011-03}.

Reconstructed photons, although not used in the main signal event selection, are selected in the regions employed in one of the alternative methods used to check the \Zjets\ background, as explained in \Sect~\ref{sec:background}. Photon candidates are required to have \pt~$ > 130$~\GeV\ and $|\eta| < 2.37$, to satisfy the tight photon shower shape and electron rejection criteria \cite{ATLAS-CONF-2012-123}, and to be isolated. 


\section{Event selection}
\label{sec:selection}
The selection of events is similar to that used in \Refe~\cite{SUSY-2013-05} and is based on the definition of one set of three overlapping signal regions (SRA) and a fourth distinct signal region (SRB). These were re-optimized for \intlumi\ of 13~\TeV\ $pp$ collisions, and the selection criteria are summarized in Table~\ref{tab:SRs}. Only events with \etmiss$>250$~\GeV\ are retained to ensure full efficiency of the  trigger. 
Jets are ordered according to decreasing \pt.
Contamination from backgrounds with high jet multiplicity, particularly \ttbar production, is suppressed by vetoing events with a fourth jet with $\pt>50$~\gev.
To discriminate against multijet background, events where \etmiss is aligned with a jet in the transverse plane are rejected by requiring $\dphimin>0.4$, where $\dphimin$ is the minimum azimuthal distance between \etmiss and the leading four jets, and by requiring $\etmiss/\meff>0.25$, where $\meff$ is defined as the scalar sum of the $\etmiss$ and the $\pt$ of the two leading jets. For signal events, no isolated charged leptons are expected in the final state, and events where a baseline electron or muon is reconstructed
are discarded.

The first set of signal regions, SRA, selects events with large \etmiss where  the two leading jets are \btagged to target models with a large mass difference between the bottom squark and the neutralino.  The main discriminating variable is  the contransverse mass (\mct) \cite{Tovey:2008ui}, which is a kinematic variable that can be used to measure the masses of pair-produced semi-invisibly decaying heavy particles. For two identical decays of heavy particles into two visible particles (or particle aggregates) $v_{1}$ and $v_{2}$, and two invisible particles, $\mct$ is defined as

\begin{equation}
      \mct^{2}(v_{1},v_{2}) = \left[ E_{\rm T}(v_{1}) + E_{\rm T}(v_{2}) \right]^2 -  \left[  \textit{{\textbf p}}_{\rm T}(v_{1})
- \textit{{\textbf p}}_{\rm T}(v_{2}) \right]^2. \nonumber
   \label{eq:mct_endpoints}
 \end{equation}

In this analysis, $v_1$ and $v_2$ are the two leading \bjets. For signal events, these correspond to  the \bjets from the squark decays and the invisible particles are the two neutralinos. 
The contransverse mass is invariant under equal and opposite boosts of the parent particles in the transverse plane. For systems of parent particles produced with small transverse boosts, \mct\ is 
bounded from above by an analytical combination of particle masses. This bound is saturated when the two 
visible objects are collinear.  For \ttbar\ events, this kinematic bound is at 135~\GeV, while for production of bottom squark 
pairs the bound is given by
$m_{\rm CT}^{\rm max} = (m^2_{\bone} - m^2_{\neut} ) / m_{\bone}$.  
The selection on \mct\ is optimized based on the bottom squark and neutralino masses considered and SRA is further divided into three overlapping regions, 
SRA250, SRA350 and SRA450, where the naming conventions reflects the minimum value allowed for \mct. Finally, a selection on the invariant mass of the two \bjets ($\mbb>200$~\GeV) is applied to further enhance the signal yield over the SM background contributions. For a signal model corresponding to 
$m^{}_{\bone}$ = 800~\GeV\ and  $m^{}_{\neut}$ = 1~\GeV, 10\%, 8\% and 5\% of the simulated signal events are retained by the 
SRA250, SRA350 and SRA450 selections, respectively.    

The second type of signal region, SRB, selects events where a bottom squark pair is produced in association with a jet from initial-state radiation (ISR). The SRB targets models with a small mass difference between the \bone and the \neut, such that a boosted bottom squark pair is needed to satisfy the trigger requirements. Hence events are selected with large \etmiss, one high-\pt\  non-\btagged\ leading jet and at least two additional \bjets. The leading jet is also required to be pointing in the direction opposite to the \etmiss\ by requiring $\dphij1 > 2.5$, where \dphij1 is  defined as the azimuthal angle between the leading jet and the \etmiss.  
For a signal model corresponding to $m^{}_{\bone}$ = 400~\GeV\ and  $m^{}_{\neut}$ = 300~\GeV, about 0.3\% of the simulated events are retained by the SRB selection.

\begin{table}[ht!]
\begin{center} 
\caption{Summary of the event selection in each signal region. The term lepton is used in the table to refer to baseline electrons and muons. 
Jets ($j_1$, $j_2$, $j_3$ and $j_4$) are labelled with an index corresponding to their decreasing order in \pt.}
\label{tab:SRs}
\begin{tabular}{@{}lcc@{}}
 \toprule
{\bf Variable }  & SRA & SRB \\
\midrule
 Event cleaning &\multicolumn{2}{c}{Common to all SR}\\ 
Lepton veto &\multicolumn{2}{c}{No $e$/$\mu$ with $\pt$ $>$ 10~\GeV\ after overlap removal}\\ 
\etmiss~[\gev]  &$> 250$  & $>400$ \\ 
$n$(jets) $\pt > 35$~\GeV & 2--4 & 3--4 \\ 
1st jet $\pt(j_1)$~[\gev]  & $> 130$  & $> 300$  \\ 
2nd jet $\pt(j_2)$~[\gev]      & $> 50$   & $> 50$   \\ 
4th jet  &  \multicolumn{2}{c}{Vetoed if $\pt(j_4)>$ 50 \gev} \\ 
$\dphimin $ & $>0.4$ & $> 0.4$ \\ 
$\dphij1 $ & - & $> 2.5$ \\ 
 {\btagging} & $j_1$ and $j_2$ & $j_2$ and ($j_3$ or $j_4$) \\
$\met/\meff$ & $> 0.25$ &  $>$ 0.25 \\ 
$ \mct $~[\gev] & $>$  250, 350, 450  & - \\ 
$m_{bb}$~[\gev] & $> 200$ & - \\ 
\bottomrule
\end{tabular}
\end{center}
\vspace{-0.5cm}
\end{table}


\section{Background estimation}
\label{sec:background}
The dominant SM background processes in the signal regions are the production of $W$ or  $Z$ bosons in association with heavy-flavour jets (referred to as \WHF and \ZHF) and the production of top quarks.  In particular, events with \ZHF production followed by the $Z\rightarrow \nu\bar{\nu}$ decay have the same signature as the signal and are the largest (irreducible) background in SRA. The background in SRB is dominated by top-quark production in events with a charged lepton in the final state that is not reconstructed, either because the lepton is a hadronically decaying $\tau$, or because the electron or muon is not identified or out of detector acceptance. 

Monte Carlo simulation is used to estimate the background yield in the signal regions, after normalizing the Monte Carlo prediction for the major backgrounds to data in control regions (CR) constructed to enhance a particular background and to be kinematically similar but orthogonal to the signal regions. The control regions are defined by explicitly requiring the presence of one or two leptons (electrons or muons)  in the final state and applying further selection criteria similar to those of the corresponding signal regions. To select events with good-quality electrons and muons the ``signal lepton'' selection is applied to them. Events with additional baseline lepton candidates are vetoed.  
For SRA, the normalizations of the top-quark pair, single-top-quark,  \Zjets\ and \Wjets\  backgrounds are estimated simultaneously by making use of four control regions, while for  
SRB two control regions are used to determine the normalization  of the top-quark pair and \Zjets\  backgrounds since other contributions are sub-dominant. A likelihood function is built as the product of Poisson probability functions, 
using as constraints  the observed and expected (from MC simulation) event yields in the control regions but not the yields in the corresponding SR~\cite{Baak:2014wma}.
The normalization factors for each of the simulated backgrounds are adjusted simultaneously via a profile likelihood fit~\cite{Cowan:2010js} 
(referred to as the ``background-only fit''). All systematic uncertainties (discussed in \Sect~\ref{sec:systematic}) are treated as nuisance parameters in the fit.  
The background normalization parameters are applied to the signal regions also taking into account correlations in the yield predictions between different regions.

Two same-flavour opposite-sign (SFOS) two-lepton (electron or muon) control regions with dilepton invariant mass near the $Z$ boson mass ($76< \mll  <106$~\GeV) and two \btagged\ jets provide data samples dominated by $Z$  boson production.  For these control regions, labelled in the following as CRzA and CRzB for SRA and SRB respectively, the \pt of the leptons is added vectorially to the \ptmissvec to mimic the expected missing transverse momentum spectrum of $Z\rightarrow \nu \bar{\nu}$ events, and is indicated in the following as $\etmisscor$ (lepton corrected). In addition, the uncorrected \etmiss of the event is required to be less than 100 (70)~\GeV\ in CRzA (CRzB) in order to further enhance  the $Z$ boson contribution. In the case of CRzA, a $\mbb>200$~\GeV\ selection is also imposed. Two control regions, CRttA and CRttB defined for SRA and SRB respectively, dominated by \ttbar production are identified by selecting events with exactly one lepton ($e,\mu$) and a  set of requirements similar to those for SRA and SRB, with the additional requirement $\mbb<200$~\GeV\ in CRttA to separate the \ttbar pair contribution from single top-quark production. To assist in the estimation of the background from \WHF and single top-quark production in SRA two further one-lepton control regions (CRwA and CRstA) are defined. These control regions  exploit kinematic features to differentiate these processes from \ttbar\ production. For CRstA a selection is applied to the minimum invariant mass of either of the \bjets and the charged lepton (\mlbmin). For CRwA only one \bjet\ is required, the selection on $\mbb$ is replaced by a selection on the invariant mass of the two leading jets ($m_{bj}$) and a further selection is applied to the event transverse mass \mt, defined as
$\mt = \sqrt{2 {p}_\mathrm{T}^{\rm lep} \etmiss - 2 \ptlep \cdot \ptmissvec} $. The definitions of the control regions are summarized in Table~\ref{tab:CRdef}.  

The contributions from diboson and $\ttbar + W/Z$ processes are minor and they are collectively called ``Others'' in the following. They are estimated from MC simulation for both the 
signal and the control regions and included in the fit procedure, and are allowed to vary within their uncertainty. The background from multijet production is estimated from data using a procedure described in detail in \Refe~\cite{SUSY-2011-20} and modified to account for the flavour of the jets. The procedure consists  of smearing the jet response in events with well-measured \etmiss\ (seed events). 
The jet response function is obtained from MC dijet events and
cross-checked in events from data where the \etmiss\ can be unambiguously
attributed to the mis-measurement of one of the jets.
The contribution from multijet production in all regions is found to be negligible.

\begin{table}[ht!]
\begin{center}
\caption{Definition of the control regions associated with SRA and SRB. Control regions are defined to study the contribution from \ZHF, \ttbar, single top-quark and \WHF\ production in SRA and the contribution from \ZHF and \ttbar in SRB. The term lepton is used in the table to refer to signal electrons and muons. Jets ($j_1$, $j_2$, $j_3$ and $j_4$)  are labelled with an index corresponding to their decreasing order in \pt. SFOS indicates the same-flavour opposite-sign two-lepton selection. }
\label{tab:CRdef}
\begin{tabular}{@{}lcccc|cc@{}}
 \toprule
   {\bf Variable} &  CRzA & CRttA & CRstA & CRwA & CRzB & CRttB\\
	    \midrule
	    Number of leptons   & 2 SFOS  & 1 & 1 & 1 & 2 SFOS & 1\\
	    1st lepton \pt  [\GeV] &  $>90$ & $>26$ & $>26$ & $>26$ & $>26$ & $>26$  \\
	    2nd lepton \pt  [\GeV] & $>20$ & - & - & - & $>20$ & - \\
	    \mll [\GeV] &   [76, 106] & -  & - & -  & [76, 106] & - \\
	    \mt  [\GeV] &   - &  - & - &  $> 30$ & - & - \\ 
            $n$(jets) $\pt > 35$~\GeV & 2--4 & 2--4 & 2--4 & 2--4   & 3--4 & 3--4 \\ 
	    1st jet $\pt(j_1)$  [\GeV]   & $> 50$  & $ > 130$ & $> 50$ & $ > 130$ & $> 50$ & $> 130$\\
	    2nd jet $\pt(j_2)$ [\GeV]  &  $> 50$  & $ > 50$ & $> 50$ & $ > 50$ & $> 50$ & $> 50$ \\
	    4th jet  &\multicolumn{4}{c|}{Vetoed if $\pt(j_4)> 50$ \gev}   & \multicolumn{2}{c}{Vetoed if $\pt(j_4)> 50$ \gev} \\
	    \btagged jets  & $j_1$ and $j_2$ & $j_1$ and $j_2$ & $j_1$ and $j_2$ & $j_1$ & \multicolumn{2}{c}{ $j_2$ and ($j_3$ or $j_4$) } \\
	    \etmiss [\GeV]   & $ < 100$ & $ > 100$ & $ > 100$ & $ > 100$ & $<70$ & $>200$ \\
	    \etmisscor  [\GeV]   & $ > 100$  & - & - & -  & $>100$ & - \\
	    \mbb~  [\GeV] &   $>200$  & $< 200$ & $ > 200$ & $ (m_{bj}) > 200 $ & - & - \\
	    $\mct$ [\GeV]   & - & $ > 150$ & $ > 150$ & $ > 150$ & - & - \\
	    \mlbmin  [\GeV] &  - &  - & $>170$  & - & - & - \\
	       $\Delta\phi(j_1,~\etmiss)$  & - & - & - & - & $ > 2.0$ & $ > 2.5$  \\
    
	    \bottomrule
	  \end{tabular}
	  \end{center}
\end{table}

\begin{table}
\begin{center}
\setlength{\tabcolsep}{0.0pc}
{\small
\caption{Fit results in the control regions associated with the SRA and SRB 
selection for an integrated luminosity of \intlumi. The results are obtained from the control regions using the background-only fit. The uncertainties include statistical, detector-related and theoretical systematic components. The individual uncertainties can be correlated and do not necessarily add  in quadrature to the total systematic uncertainty. The pure MC estimate is used for backgrounds for which a dedicated CR is not defined, e.g. for smaller backgrounds (indicated as ``Other'') and for single top-quark and \Wjets\ production in CRzB and CRttB. A dash indicates a negligible background.}
\label{tab:bkgfit}
\begin{tabular*}{\textwidth}{@{\extracolsep{\fill}}lcccccc}
\noalign{\smallskip}\hline\noalign{\smallskip}
{\bf Control Region}           & CRzA            & CRwA            & CRttA            & CRstA            & CRzB            & CRttB             \\[-0.05cm]
\noalign{\smallskip}\hline\noalign{\smallskip}

Observed           & $78$              & $543$              & $260$              & $56$              & $59$              & $188$                    \\
\noalign{\smallskip}\hline\noalign{\smallskip}
Total background (fit)         & $\NumUnc{78}{9}$          & $\NumUnc{543}{23}$          & $\NumUnc{260}{16}$          & $\NumUnc{56}{7}$          & $\NumUnc{59}{8}$          & $\NumUnc{188}{14}$              \\
\noalign{\smallskip}\hline\noalign{\smallskip}
         \ttbar          & $9.0 \pm 1.6$          & $\NumUnc{153}{26}$\           & $\NumUnc{181}{23}$          & $\NumUnc{11.1}{2.1}$          & $\NumUnc{14.6}{2.0}$          & $\NumUnc{156}{15}$              \\
         Single top       & $0.8 \pm 0.4$          & $50 \pm 23$          & $27 \pm 13$          & $23 \pm 10$          & $0.42 \pm 0.07$          & $\NumUnc{16.6}{2.0}$              \\
         $W$+jets         & -          & $\NumUnc{327}{43}$          & $45 \pm 14$          & $\NumUnc{20}{6}$          & -          & $\NumUnc{13}{5}$              \\
         $Z$+jets          & $\NumUnc{68}{9}$          & $3.8 \pm 0.6$          & $1.4 \pm 0.2$          & $0.9 \pm 0.2$          & $\NumUnc{42}{8}$          & $0.3 \pm 0.1$              \\
         ``Other''         & $0.9 \pm 0.1$          & $8.1 \pm 1.1$          & $5.8 \pm 0.7$          & $0.6 \pm 0.1$          & $1.6 \pm 0.4$          & $2.3 \pm 0.2$              \\
 \noalign{\smallskip}\hline\noalign{\smallskip}
Total background (MC exp.)             & $61$          & $503$          & $267$          & $57 $          & $46$          & $191 $              \\
\noalign{\smallskip}\hline\noalign{\smallskip}
         \ttbar         & $9.4 $          & $161 $          & $190$          & $12 $          & $15$          & $159 $              \\
        Single top          & $1.1 $          & $60 $          & $33$          & $27 $          & $0.4 $          & $16 $              \\
        $W$+jets          & -          & $270 $          & $37 $          & $17 $          & -          & $12.9 $              \\
        $Z$+jets          & $50 $          & $2.8$          & $1.0 $          & 0.7         & $29 $          & $0.2$              \\
        ``Other''         & $0.9 $          & $8.1 $          & $5.8$          & $0.6 $          & $1.6$         & $2.3 $              \\
\noalign{\smallskip}\hline\noalign{\smallskip}

\end{tabular*}
}
\end{center}
\end{table}

\begin{figure}[htbp]
   \centering
   \begin{center}
   \includegraphics[width=0.49\textwidth, angle=0]{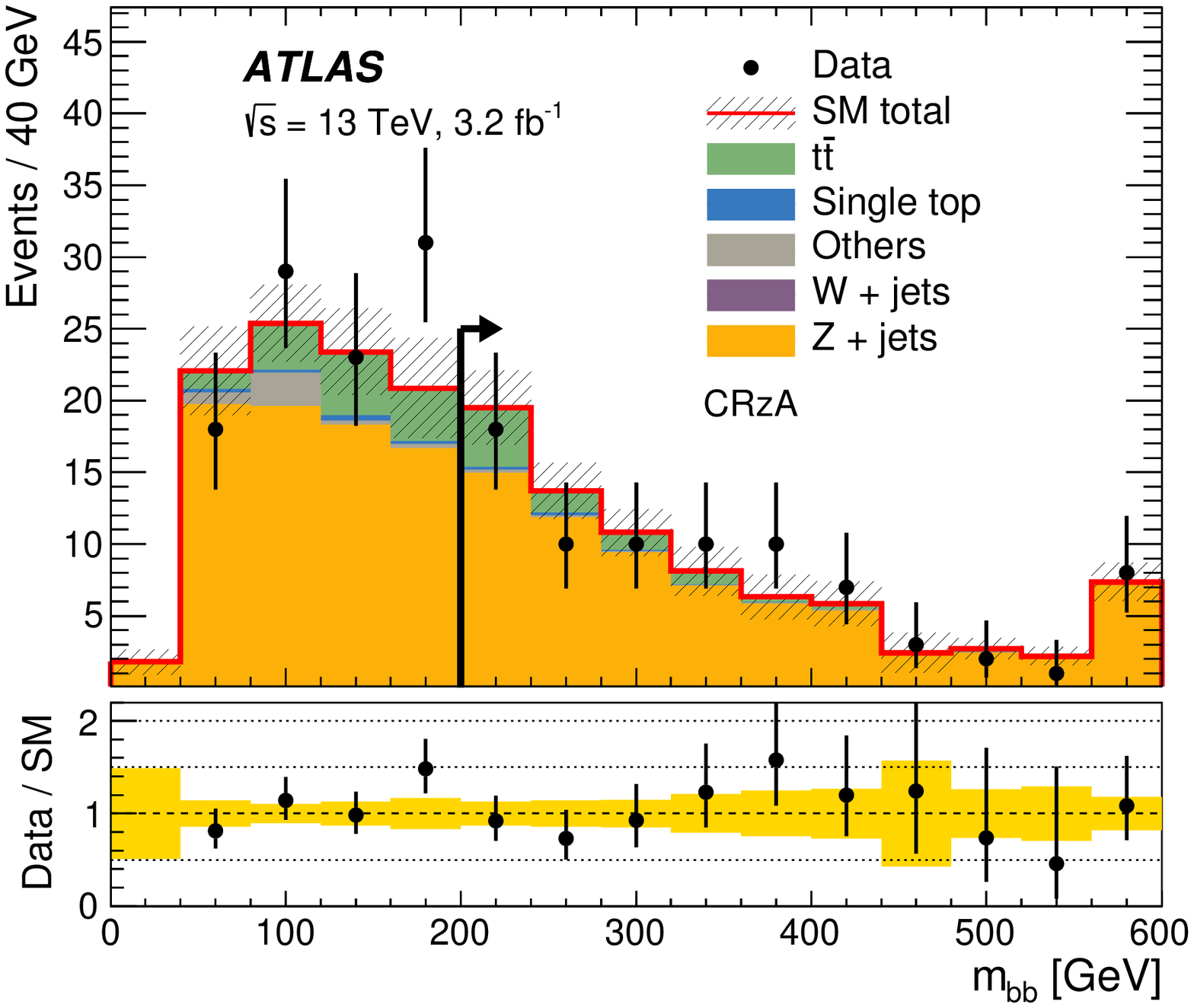}
   \includegraphics[width=0.49\textwidth, angle=0]{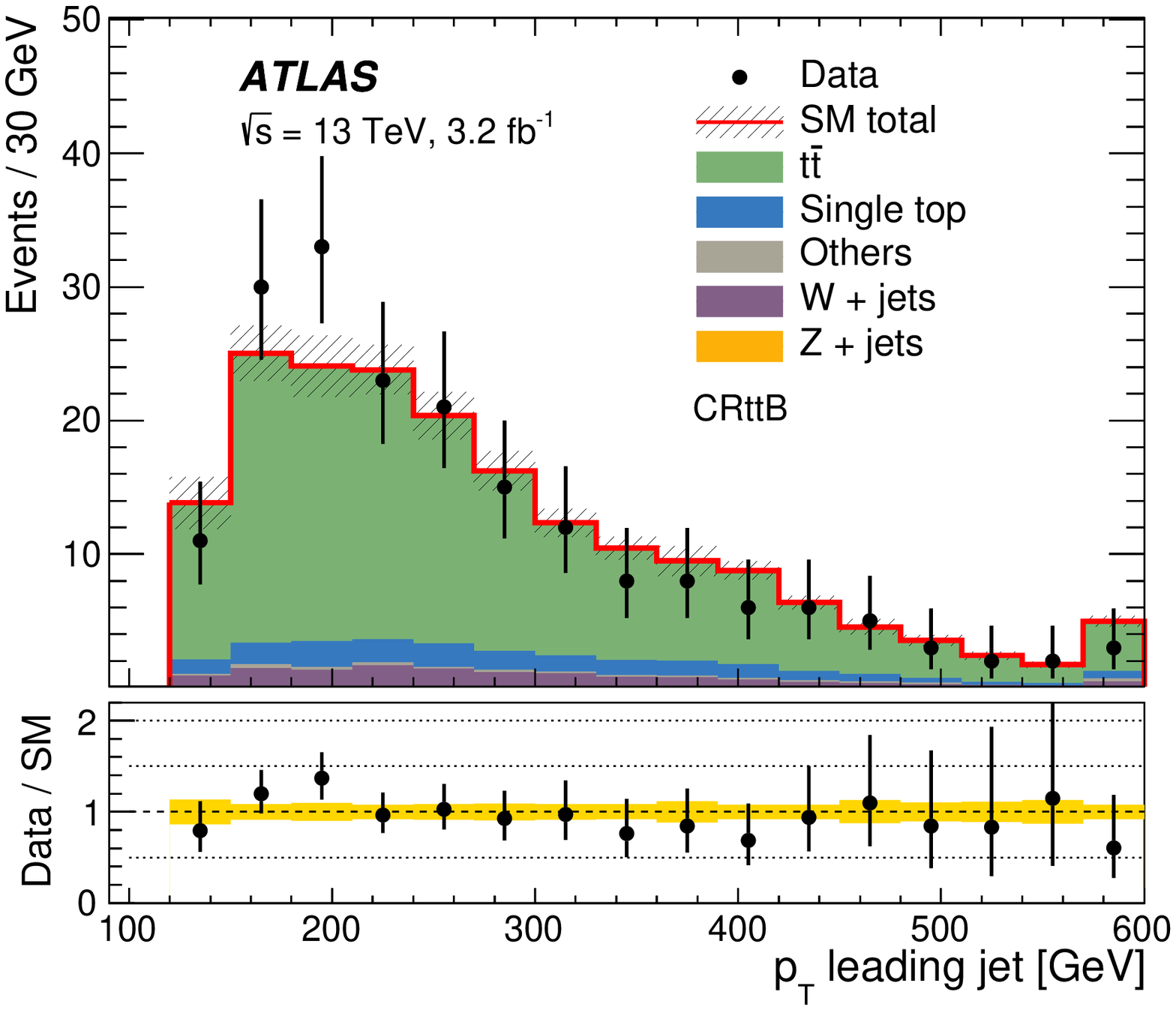}
   \end{center}
   \caption{Left: \mbb\ distribution in CRzA before the final selection $\mbb>200$~\GeV\ is applied (indicated by the arrow). Right: \pt of the leading jet in CRttB. In both distributions the MC normalization is rescaled using the results from the background-only fit, showing good agreement between data and the predicted SM shapes. The shaded band includes statistical and  detector-related systematic uncertainties as detailed in \Sect~\ref{sec:systematic} and the last bin includes overflows.}
   \label{fig:CRplots}
\end{figure}

The results of the background-only fit  are shown in Table~\ref{tab:bkgfit}, where the contribution from individual backgrounds is shown separately as a purely MC-based prediction and with the rescaling from the fit procedure. 
The \mbb\ distribution in CRzA and the \pt of the leading jet in CRttB are shown in \fig~\ref{fig:CRplots} after the backgrounds were rescaled as a result of the background-only fit, showing good agreement in the shape of the distributions in the two control regions used to estimate the dominant backgrounds in SRA and SRB.

The full background estimation procedure is validated by comparing the background predictions and the shapes of the distributions of the key analysis variables from the fit results to those observed in dedicated validation regions. They are defined to be mutually exclusive and kinematically similar to the signal regions, with low potential contamination from signal. For SRA, two validation regions are defined by using the same definition as SRA250, and inverting the selections on \mbb\  and \mct. The SM predictions are found to overestimate the data by about one standard deviation. For SRB, a validation region is defined by selecting events with 250~$<\etmiss <$~300~\GeV\ and good agreement is observed between data and predictions. 

As a further validation, two alternative methods are used to estimate the \ZHF contribution. 
The first method exploits the similarity of the \Zjets\ and $\gamma+$jets processes
\cite{SUSY-2011-20}. For \pt\ of the photon significantly larger than the mass of the $Z$ boson, the kinematics of $\gamma+$jets events strongly resemble those of \Zjets\ events. The event yields are measured in control regions identical to the SRA and SRB, with the \etmiss-based selections replaced by selections on the \pt\ of the photon vectorially added to the \ptmissvec. 
The yields are then propagated to the actual SRA and SRB using a reweighting factor derived using the MC simulation. 
This factor takes into account the different kinematics of the two processes and residual effects arising from 
the acceptance and reconstruction efficiency for photons.

In the second alternative method, applied to SRA only where the \ZHF contribution is dominant, the MC simulation is used to verify that the shape of the \mct\ distribution for events with no \btagged\ jets is compatible with the shape of the \mct\ distribution for  events where two \btagged\ jets are present. A  new highly populated \Zjets\ CR is defined selecting $Z\rightarrow \ell \ell$ events with no $b-$tagged jets. The \mct\ distribution in this CR is constructed using the two leading jets and is used to estimate the shape of the \mct\ distribution in the SRA, whilst the normalization in SRA is rescaled based on the ratio in data of $Z \rightarrow \ell \ell$ events with no \btagged\ jets to events with two \btagged\ jets. Additional MC-based corrections are applied to take into account the two-lepton selection in this CR.
 
The two alternative methods are in agreement within uncertainties with the estimates obtained with the profile likelihood fit to the control regions (Table~\ref{tab:Zjets_A_summary}). 
Experimental and theoretical systematic uncertainties  in the nominal and alternative method estimates are taken into account (see \Sect~\ref{sec:systematic}). The difference between the alternative methods and the background-only fit is taken into account as an additional systematic uncertainty in the final \ZHF yields.  

\begin{table}[htbp]
\begin{center}
\setlength{\tabcolsep}{0.0pc}
\caption{Estimated \Zjets yields in the signal regions as obtained using the default and the two alternative methods. The errors include all the uncertainty sources discussed in \Sect~\ref{sec:systematic}. The ``MC-based post-fit'' uncertainty  does not include the additional uncertainty to account for the difference between the three methods.}
\label{tab:Zjets_A_summary}
{\small
  \begin{tabular*}{\textwidth}{@{\extracolsep{\fill}}lrrrr}
  \noalign{\smallskip}\hline\noalign{\smallskip}
  {\bf Method \textbackslash~Region}    & SRA250                   & SRA350                  & SRA450         & SRB     \\[-0.05cm]
     \noalign{\smallskip}\hline\noalign{\smallskip}
  Nominal MC-based post-fit   &  $22 \pm  4$       &  $5.0 \pm  0.9 $      &  $1.3  \pm  0.3$  & $4.1 \pm 0.9 $ \\
  $Z$+jets from $\gamma+$jets events & $18 \pm 5$           & $3.7 \pm 1.5$      & $1.8 \pm 1.0$         &  $2.2 \pm 1.0$              \\
  $Z$+jets from non $b$-tagged $Z$ events & $18 \pm 6$           & $4.3 \pm 1.6$      & $1.3 \pm 0.5$         &  Not Applicable        \\
    \noalign{\smallskip}\hline\noalign{\smallskip}
    \end{tabular*}
     }
     \end{center}
     \end{table}


\section{Systematic uncertainties}
\label{sec:systematic}
Several sources of experimental and theoretical systematic uncertainty are considered in this analysis. 
Their impact is reduced through the normalization of the dominant backgrounds in the control regions with kinematic selections resembling those of the corresponding signal region (see \Sect~\ref{sec:background}). Uncertainties due to the limited number of events in the CRs are also taken into account in the fit. The individual contributions are outlined in Table~\ref{tab:syst_summary}

\begin{table}[htbp]
\begin{center}
\setlength{\tabcolsep}{0.0pc}
\caption{Summary of the dominant experimental and theoretical uncertainties for each signal region. Uncertainties are quoted as 
relative to the total uncertainty, with a range indicated for the three SRAs. For theoretical modelling, uncertainties per dominant SM background process are quoted. 
The individual uncertainties can be correlated, and do not necessarily add in quadrature to the total background uncertainty.}
\label{tab:syst_summary}
{\small                                                                                                                                                                                           
  \begin{tabular*}{\textwidth}{@{\extracolsep{\fill}}lcc}
    \noalign{\smallskip}\hline\noalign{\smallskip}
  {\bf Source \textbackslash~Region}    &     SRA    & SRB     \\[-0.05cm]
     \noalign{\smallskip}\hline\noalign{\smallskip}
  Experimental uncertainty  &  & \\ 
  \midrule
  JES  & 15--30\% & 25\% \\
  JER  & 20--35\% & $<$10\% \\ 
  $b$-tagging & 25--45\% & 15\% \\ 
  \midrule
  Theoretical modelling uncertainty  &  & \\ 
  \midrule 
  $Z$+jets & 25--35\% & 12\% \\ 
  $W$+jets & 20--22\% & 27\% \\ 
  Top production & 15--20\% & 70\% \\ 
    \noalign{\smallskip}\hline\noalign{\smallskip}
    \end{tabular*}
     }
     \end{center}
     \end{table}

The dominant detector-related systematic effects are due to the uncertainties in the jet energy scale (JES)~\cite{ATL-PHYS-PUB-2015-015}  and resolution (JER)~\cite{ATLAS-CONF-2015-037}, and in the $b$-tagging efficiency and mis-tagging rates. The JES and JER uncertainties are estimated from 13~\TeV\ data, while the uncertainties related to \btagging are estimated  from 8~\TeV\ data 
and extrapolated to 13~\TeV\ and to the Run-2 detector conditions.  The uncertainties associated with lepton and photon reconstruction and energy measurements are also considered but have a small impact on the final results. Lepton, photon and jet-related uncertainties are propagated to the calculation of the \etmiss, and additional uncertainties are included in the energy scale and resolution of the soft term. 
The overall experimental uncertainty in the SM background estimate is found to be around 20\% for the SRAs and 15\% for the SRB. 
 
Uncertainties in the modelling of the SM background processes from MC simulation and their theoretical cross-section uncertainties are also taken into account. The dominant uncertainty arises from $Z$+jets MC modelling for SRA and $\ttbar$ modelling for SRB. The $Z$+jets (as well as $W$+jets) modelling uncertainties are evaluated using alternative samples generated with different renormalization and factorization scales, merging (CKKW-L) and resummation scales. 
An additional one-sided uncertainty in the \Zjets\ estimate is taken as the largest deviation between the nominal background-only fit result and each of the alternative data-driven estimates described in \Sect~\ref{sec:background}. This results in an additional 25\%, 25\% and 40\% one-sided uncertainty in SRA250, SRA350 and SRA450, respectively. 
Finally, a 40\% uncertainty~\cite{STDM-2011-21} is assigned to the heavy-flavour jet content in $W$+jets, estimated from MC simulation in SRB. For SRA, the uncertainty accounts for the different requirements on \bjets between CRwA and the signal region.  

Uncertainties in the modelling of the top-quark pair and single-top-quark ($Wt$) backgrounds are sub-dominant in SRA and dominant in SRB. They are computed as the difference between the predictions from nominal samples and those of additional samples differing in generator or parameter settings. Hadronization and PS uncertainties are estimated using samples generated with \powhegbox v2 and showered by \herwigpp v2.7.1~\cite{Bahr:2008pv} with the UEEE5 underlying event tune. Uncertainties related to initial- and final-state radiation modelling, tune and (for \ttbar~only) choice of $h^{}_{\rm damp}$ parameter in \powhegbox v2 are evaluated using alternative settings of the generators. Finally, an alternative generator \madgraph with showering by \herwigpp v2.7.1 is used to estimate the generator uncertainties. Uncertainties in smaller backgrounds such as diboson and $ttV$ are also estimated by comparisons of the nominal sample with alternative samples differing in generator or parameter settings (\powheg v2 with showering by \pythia~8.210 for diboson, renormalization and factorization scale and A14 tune variations for $ttV$) and are found to be negligible. The cross-sections used to normalize the MC yields to the highest order available are varied according to the scale uncertainty of the theoretical calculation, i.e. 5\% for $W$, $Z$ boson and top-quark pair production, 6\% for diboson, 13\% and 12\% for $ttW$ and $ttZ$, respectively.   

For the SUSY signal processes, both the experimental and theoretical uncertainties in the expected signal yield are considered. Experimental uncertainties, which are found to be between 20\% and 25\% across the $\bone$--$\neut$ mass plane for all SRs, are largely dominated by the uncertainty in the $b$-tagging efficiency in SRA, while JER and $b$-tagging uncertainties are dominant in SRB with equal contributions. Theoretical uncertainties in the NLO+NLL cross-section are calculated  for each SUSY signal scenario and are dominated by the uncertainties in the renormalization and factorization scales, followed by the uncertainty in the PDF. 
They vary between 15\% and 20\% for bottom squark masses in the range between 400~\GeV\ and 900~\GeV.  
Additional uncertainties in the modelling of initial-state radiation in SUSY signal MC samples are taken into account and contribute up to 5\%.


\section{Results and interpretation}
\label{sec:results}
Table~\ref{tab:fit_SRA_SRB} reports the observed number of events and the SM predictions after the background-only fit 
for each signal region. The largest background contribution in SRA arises from $Z\rightarrow \nu\bar{\nu}$ produced in 
association with $b$-quarks whilst top-quark pair production dominates SM predictions for SRB.  The background-only fit results 
are compared to the pre-fit predictions based on MC simulation. 
\begin{table}
\begin{center}
\setlength{\tabcolsep}{0.0pc}
{\small
\caption{Fit results in all signal regions for an integrated luminosity of \intlumi.
The results are obtained from the control regions. The background normalization parameters obtained with the background-only fit are applied to the SRs. 
The individual uncertainties, including detector-related and theoretical systematic components, are symmetrized and can be correlated and do 
not necessarily add  in quadrature to the total systematic uncertainty. }
\label{tab:fit_SRA_SRB}
\begin{tabular*}{\textwidth}{@{\extracolsep{\fill}}lcccc}
\noalign{\smallskip}\hline\noalign{\smallskip}
{\bf Signal region}         & SRA250                    & SRA350                   & SRA450         &  SRB     \\[-0.05cm]

\noalign{\smallskip}\hline\noalign{\smallskip}
Observed           & $23$              & $6$              & $1$                  &      $6$   \\
\noalign{\smallskip}\hline\noalign{\smallskip}
Total background (fit)         & $\NumUnc{29}{5}$          & $7.0 \pm 1.2$          & $1.8\pm 0.4$            & $\NumUnc{12.0}{2.5}$  \\
\noalign{\smallskip}\hline\noalign{\smallskip}
        \ttbar         &  $\NumUnc{1.0}{0.4}$          & $0.17 \pm 0.08$          & $0.04 \pm 0.02$             & $5.5 \pm 2.0$ \\             
        Single top         & $1.8 \pm 1.0$          & $0.53 \pm 0.30$          & $0.13 \pm 0.07$      & $1.0 \pm 0.4$ \\                
        $W$+jets         & $4.4 \pm 1.3$          & $1.2 \pm 0.4$          & $0.30 \pm 0.10$             & $1.1 \pm 0.6$     \\         
        $Z$+jets         &  $\makebox[2ex]{\hfill 22} \pm \makebox[2ex]{4\hfill}$          & $5.0 \pm 1.1$          & $1.3 \pm 0.4$            & $4.1 \pm 1.3$            \\
        ``Other''         & $0.45 \pm 0.06$          & $0.14 \pm 0.04$          & $0.04 \pm 0.04$             & $0.3 \pm 0.1$            \\
 \noalign{\smallskip}\hline\noalign{\smallskip}                                                                 
Total background (MC exp.)              & $23$          & $5.6$          & $1.5 $         &     $11$    \\
\noalign{\smallskip}\hline\noalign{\smallskip}                                                                  
         \ttbar         & $1.1$          & $0.18$          & $0.04 $             & $5.6$ \\
         Single top         & $2.2$          & $0.6$          & $0.15$      & $1.0$ \\   
         $W$+jets         & $3.6 $          & $1.0 $          & $0.25$             & $1.1$                \\
         $Z$+jets         & $16 $          & $3.7 $          & $1.0$            & $2.8$                 \\
         ``Other''         & $0.45$          & $0.14$          & $0.04 $             & $0.3$                \\
\noalign{\smallskip}\hline\noalign{\smallskip}

\end{tabular*}
}
\end{center}
\end{table}

\begin{figure}[htbp]
   \centering
   \begin{center}
   \includegraphics[width=0.49\textwidth, angle=0]{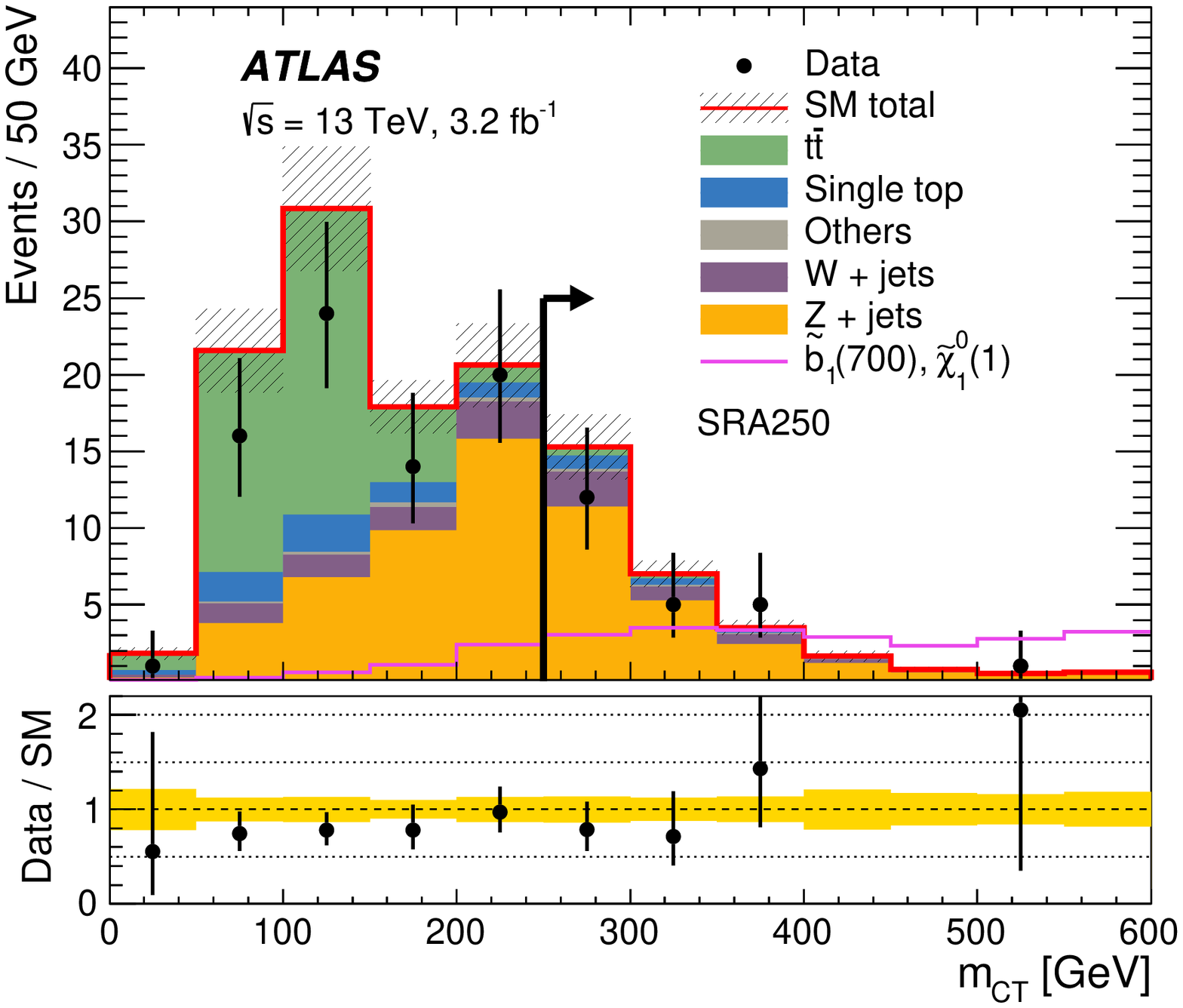} 
   \includegraphics[width=0.49\textwidth, angle=0]{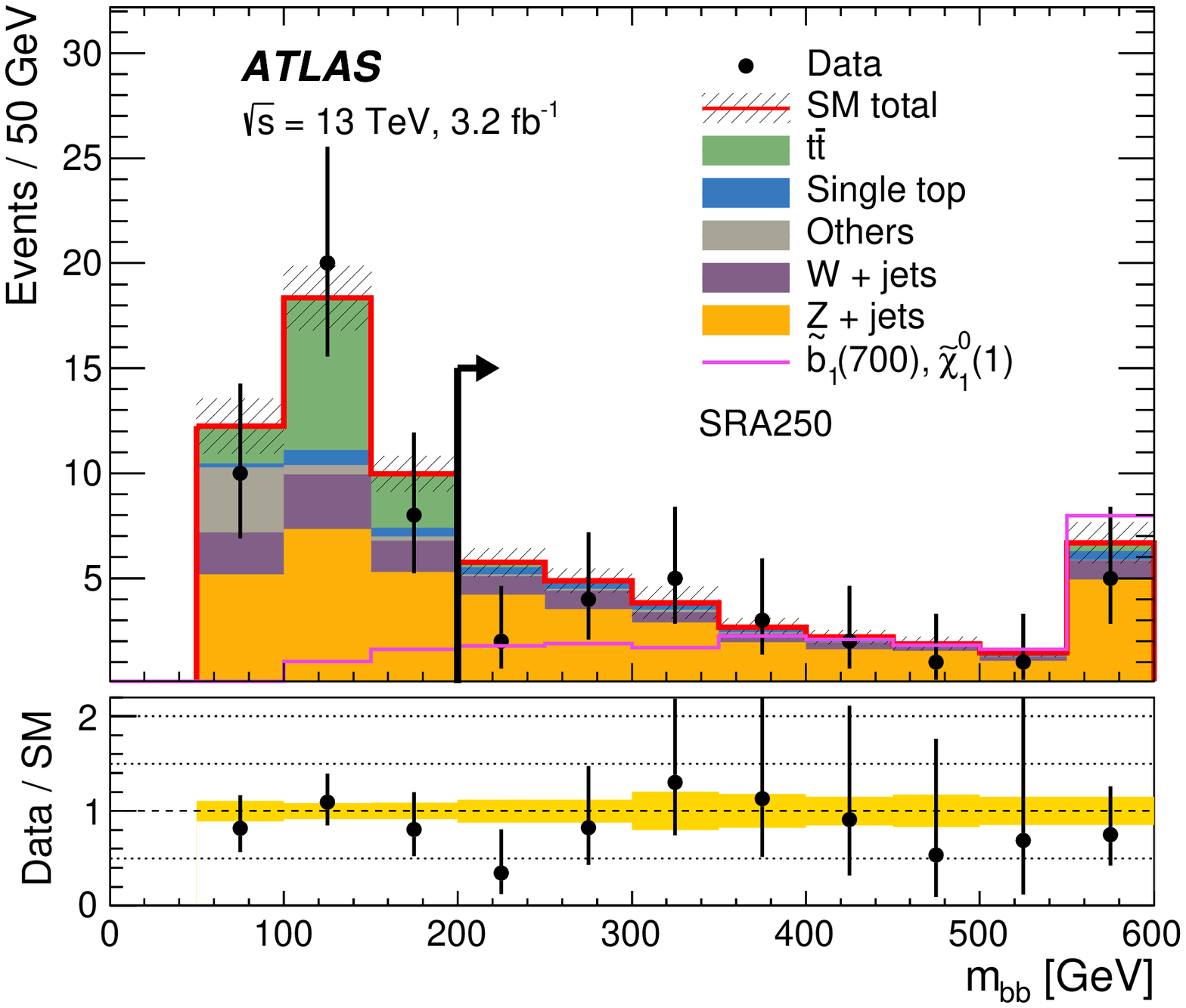} 
   \end{center}
   \caption{Left: \mct distribution in SRA250 with all the selection criteria applied except the \mct threshold. Right: 
\mbb~ distribution in SRA250 with all selection criteria applied except the \mbb\ requirement. The arrows indicate the final applied selection. The shaded band includes statistical and  
detector-related systematic uncertainties. The SM backgrounds are normalized to the values determined 
in the fit. For illustration the distributions expected for one signal model with bottom squark and neutralino masses of 700 and 1~\GeV, respectively, 
are overlaid. The last bin includes overflows. }
   \label{fig:SRA250_mctMbb}
\end{figure}

\begin{figure}[htbp]
   \centering
   \begin{center}
   \includegraphics[width=0.49\textwidth, angle=0]{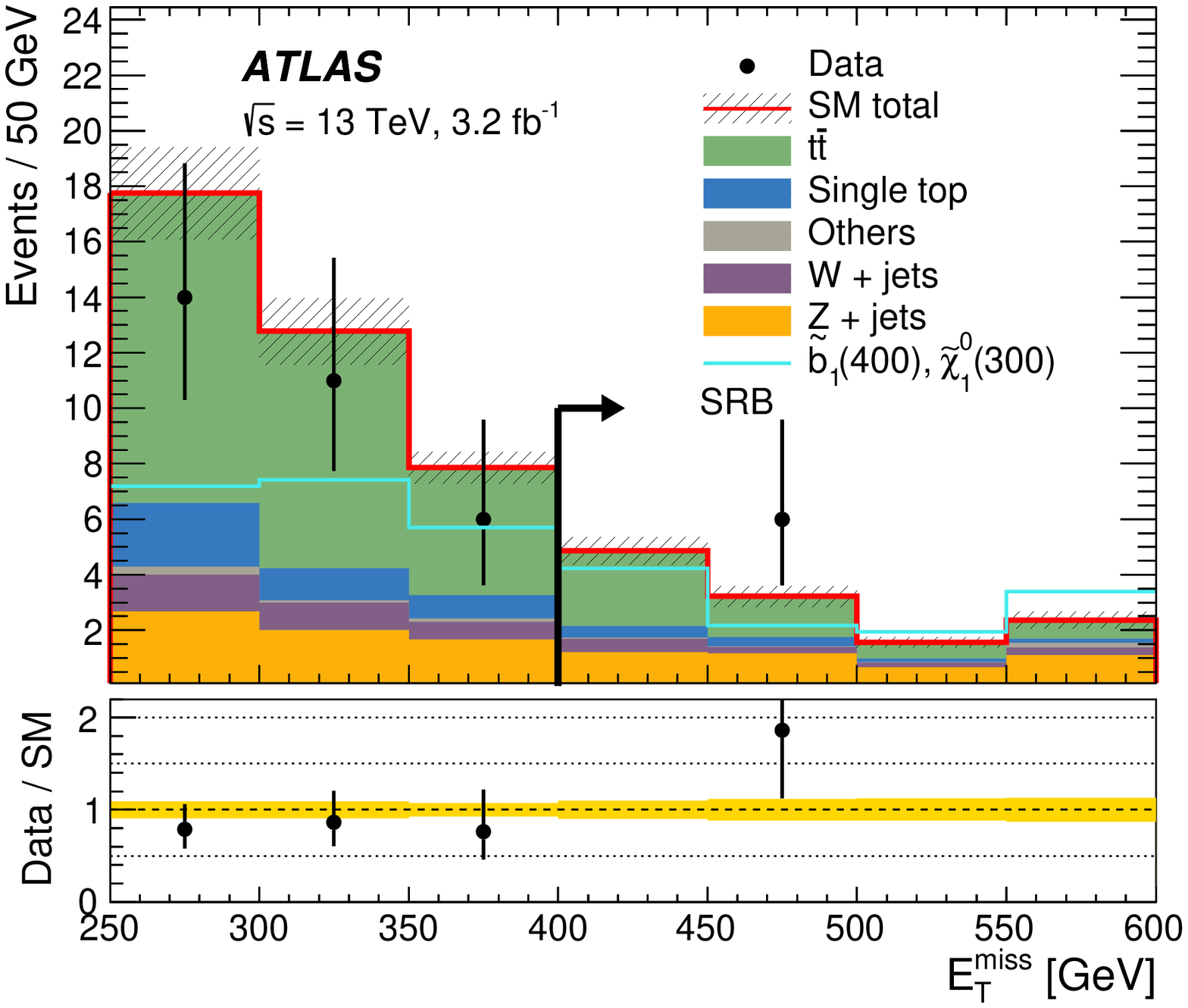} 
   \includegraphics[width=0.49\textwidth, angle=0]{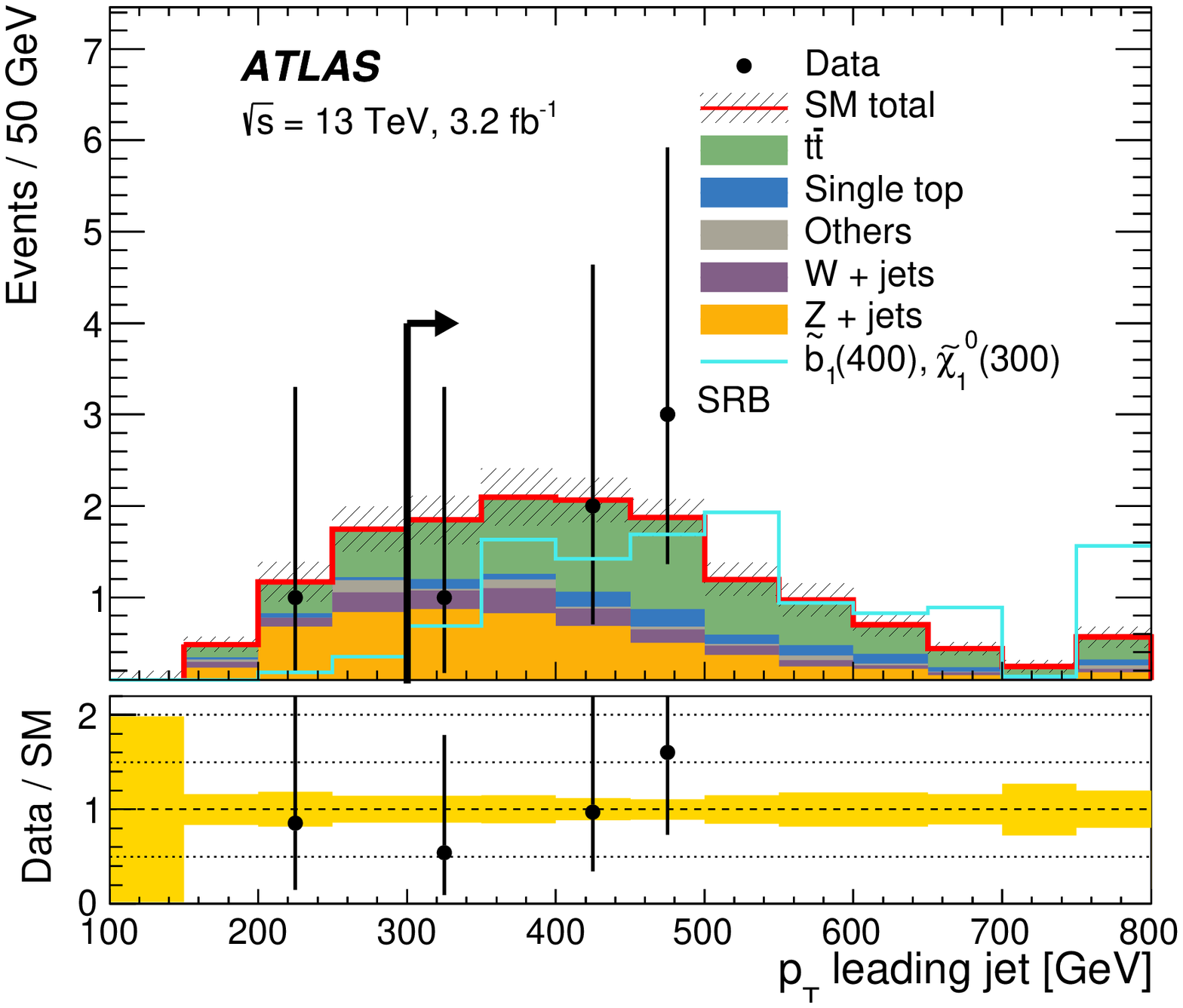} 
   \end{center}
   \caption{Left: \etmiss distribution in SRB with all the selection criteria applied except the \etmiss threshold. Right:
Leading jet \pT\ distribution in SRB with all the selection criteria applied except the selection on the leading jet \pt\ itself. The arrows indicate the final selection applied in SRB. The shaded band includes statistical and 
detector-related systematic uncertainties. The SM backgrounds are normalized to the values determined
in the fit. For illustration the distributions expected for one signal model with bottom squark and neutralino masses of 400 and 300~\GeV, respectively, 
are overlaid. The last bin includes overflows.}
   \label{fig:SRB_mctETmiss}
\end{figure}

Figures~\ref{fig:SRA250_mctMbb} and \ref{fig:SRB_mctETmiss} show the comparison between the observed data and  
the SM predictions for some relevant kinematic distributions in SRA250 and SRB, respectively, prior to 
the selection on the variable shown. 
For illustrative purposes, the distributions expected for a scenario with bottom squark and neutralino masses 
of 700~\GeV\ (400~\GeV) and 1~\GeV\ (300~\GeV), respectively, are shown for SRA250 (SRB). 
No excess above the expected Standard Model background yield is observed. 
The results are translated into upper limits on contributions from new physics beyond the SM (BSM) 
for each signal region.
The \cls~method~\cite{Junk:1999kv,Read:2002hq} is used to derive the confidence level 
of the exclusion; signal models with a \cls~value below 0.05 are said to be excluded at 95\% CL. 
The profile-likelihood-ratio test statistic is used to exclude the signal-plus-background hypothesis for
specific signal models. 
When normalized by the integrated luminosity of the data sample, results can be interpreted 
as corresponding upper limits on the visible cross-section, 
$\sigma^{}_{\rm vis}$, defined as the product of the BSM production cross-section, the acceptance and the 
selection efficiency of a BSM signal. Table~\ref{tab:upper} summarizes the observed ($S^{95}_\mathrm{obs}$) 
and expected ($S^{95}_\mathrm{exp}$) 95\% CL upper limits on the number of BSM events and on $\sigma^{}_{\rm vis}$.
\begin{table}
\begin{center}
\caption[Breakdown of upper limits.]{
Left to right: 95\% CL upper limits on the visible cross-section
($\langle\epsilon A \sigma\rangle_{\rm obs}^{95}$) and on the number of
signal events ($S_{\rm obs}^{95}$ ).  The third column
($S_{\rm exp}^{95}$) shows the 95\% CL upper limit on the number of
signal events, given the expected number (and $\pm 1\sigma$
variations of the expectation) of background events.
\label{tab:upper}}
\begin{tabular}{@{}lccc@{}}
\toprule
{\bf Signal channel}                        & $\langle\epsilon A {\rm \sigma}\rangle_{\rm obs}^{95}$[fb]  &  $S_{\rm obs}^{95}$  & $S_{\rm exp}^{95}$   \\
\midrule
SRA250   & $3.42$ &  $11.0$  & ${13.8}^{+6.0}_{-3.2}$   \\[0.1cm] 
SRA350   & $1.93$ &  $6.2$  & ${6.6}^{+3.1}_{-1.1}$    \\[0.1cm] 
SRA450   & $1.23$ &  $3.9$  & ${4.1}^{+1.9}_{-0.6}$    \\[0.1cm] 
SRB   & $1.89$ &  $6.1$  & ${8.7}^{+3.1}_{-2.5}$    \\[0.1cm] 
\bottomrule
\end{tabular}
\end{center}
\end{table}

\begin{figure}[t]
   \centering
   \begin{center}
   \includegraphics[width=0.65\textwidth, angle=0]{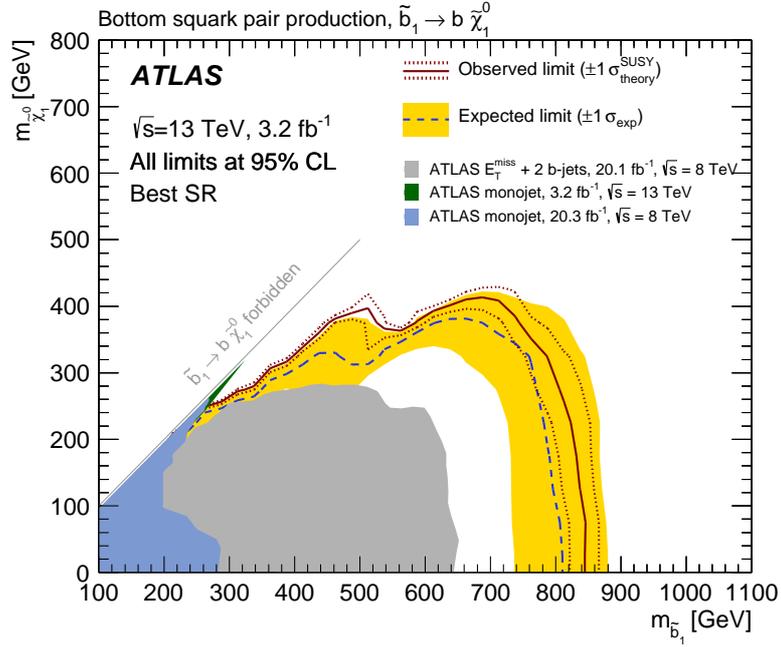}  
   \end{center}
   \caption{Observed and expected exclusion limits at 95\% CL, as well as $\pm 1 \sigma$ variation of the expected limit,  
in the $\bone$--$\neut$ mass plane. The SR with the best expected sensitivity is adopted for each point of the parameter space. 
The yellow band around the expected limit (dashed line) shows the impact of the 
experimental and SM background theoretical uncertainties. The dotted lines show the impact on the observed 
limit of the variation of the nominal signal cross-section by $\pm 1 \sigma$ of its theoretical uncertainties. 
The exclusion limits from the Run-1 ATLAS searches~\cite{SUSY-2013-21,SUSY-2013-05} and from the 13~\TeV\ monojet search~\cite{Aaboud:2016tnv} 
 are also superimposed. The latter limit is only published for values of  $m_{\bone}- m_{\neut}=$ 5 and 20~\GeV.} 
   \label{fig:limit}
\end{figure}

Exclusion limits are obtained assuming a specific SUSY particle mass hierarchy such that the lightest 
bottom squark decays exclusively via $\bone \rightarrow b \neut$. 
In this case, the fit procedure takes into account not only correlations in the yield predictions between 
control and signal regions due to common background normalization parameters and systematic uncertainties but also  
contamination in the control regions from SUSY signal events (found to be generally negligible).  
The experimental systematic uncertainties in the signal are taken into account for this calculation 
and are assumed to be fully correlated with those in the SM background. 
Figure~\ref{fig:limit} shows the observed (solid line) and expected (dashed line) 
exclusion contours at 95\% CL in the $\bone$--$\neut$ mass plane.  

At each point of the parameter space, the SR with the best expected sensitivity is adopted. 
Sensitivity to scenarios with the largest mass difference between the $\bone$ and the $\neut$~is achieved 
with the most stringent \mct~threshold (SRA450). Sensitivity to scenarios with smaller mass differences 
is achieved predominantly with searches based on the presence of a high-\pt\ ISR jet, as in the dedicated SRB selection and the search described in \Refe~\cite{SUSY-2013-21}.  
Bottom squark masses up to 800 (840)~\GeV\ are excluded for $\neut$~masses below 360 (100)~\GeV. 
Differences in mass above 100~\GeV\ between $\bone$ and $\neut$ are excluded up to $\bone$ masses of 500~\GeV. 
The expected exclusion constraints are about 20~\GeV\ lower than the observation for high bottom squark masses and about 60~\GeV\ lower for 
SUSY models with small $\bone$--$\neut$ mass splitting. 
The current results significantly extend the $\sqrt{s} = 8$~\TeV\ limits~\cite{SUSY-2013-21,SUSY-2013-05} despite the lower
integrated luminosity mostly because of the increase in centre-of-mass energy of the LHC. 
Furthermore, the sensitivity of the analysis benefits from the advanced algorithms adopted to identify $b$-jets 
and use of information from the newly installed IBL pixel layer in the Run-2 ATLAS detector, as well as from 
improved techniques to estimate SM background contributions and their systematic uncertainties.


\FloatBarrier

\section{Conclusion}
\label{sec:conclusion}
In summary, the results of a search for bottom squark pair production are reported. The analysis uses \intlumi~of $pp$ 
collisions at $\sqrt{s}=13$~\TeV\ collected by the ATLAS experiment at the Large Hadron Collider in 2015.
Bottom squarks are searched for in events containing large missing transverse momentum and up to four jets, exactly two 
of which are identified as \bjets. 
No excess above the expected Standard Model background yield is found. Exclusion limits at 95\% confidence 
level are placed on the visible cross-section and on the mass of the bottom squark in phenomenological supersymmetric 
$R$-parity-conserving models in which the $\bone$ is the lightest squark and is assumed to decay exclusively 
via $\bone \rightarrow b \neut$, where \neut~is the lightest neutralino. 
Bottom squark masses up to 800~\GeV\ are excluded for $\neut$ masses below 360~\GeV\ (840~\GeV\ for $\neut$ masses below 100~\GeV) 
whilst differences in mass above 100~\GeV\ between $\bone$ and $\neut$ are excluded up to $\bone$ masses of 500~\GeV. 
The results significantly extend the constraints on bottom squark masses with respect to Run-1 searches.


\section*{Acknowledgements}


We thank CERN for the very successful operation of the LHC, as well as the
support staff from our institutions without whom ATLAS could not be
operated efficiently.

We acknowledge the support of ANPCyT, Argentina; YerPhI, Armenia; ARC, Australia; BMWFW and FWF, Austria; ANAS, Azerbaijan; SSTC, Belarus; CNPq and FAPESP, Brazil; NSERC, NRC and CFI, Canada; CERN; CONICYT, Chile; CAS, MOST and NSFC, China; COLCIENCIAS, Colombia; MSMT CR, MPO CR and VSC CR, Czech Republic; DNRF and DNSRC, Denmark; IN2P3-CNRS, CEA-DSM/IRFU, France; GNSF, Georgia; BMBF, HGF, and MPG, Germany; GSRT, Greece; RGC, Hong Kong SAR, China; ISF, I-CORE and Benoziyo Center, Israel; INFN, Italy; MEXT and JSPS, Japan; CNRST, Morocco; FOM and NWO, Netherlands; RCN, Norway; MNiSW and NCN, Poland; FCT, Portugal; MNE/IFA, Romania; MES of Russia and NRC KI, Russian Federation; JINR; MESTD, Serbia; MSSR, Slovakia; ARRS and MIZ\v{S}, Slovenia; DST/NRF, South Africa; MINECO, Spain; SRC and Wallenberg Foundation, Sweden; SERI, SNSF and Cantons of Bern and Geneva, Switzerland; MOST, Taiwan; TAEK, Turkey; STFC, United Kingdom; DOE and NSF, United States of America. In addition, individual groups and members have received support from BCKDF, the Canada Council, CANARIE, CRC, Compute Canada, FQRNT, and the Ontario Innovation Trust, Canada; EPLANET, ERC, FP7, Horizon 2020 and Marie Sk{\l}odowska-Curie Actions, European Union; Investissements d'Avenir Labex and Idex, ANR, R{\'e}gion Auvergne and Fondation Partager le Savoir, France; DFG and AvH Foundation, Germany; Herakleitos, Thales and Aristeia programmes co-financed by EU-ESF and the Greek NSRF; BSF, GIF and Minerva, Israel; BRF, Norway; Generalitat de Catalunya, Generalitat Valenciana, Spain; the Royal Society and Leverhulme Trust, United Kingdom.

The crucial computing support from all WLCG partners is acknowledged gratefully, in particular from CERN, the ATLAS Tier-1 facilities at TRIUMF (Canada), NDGF (Denmark, Norway, Sweden), CC-IN2P3 (France), KIT/GridKA (Germany), INFN-CNAF (Italy), NL-T1 (Netherlands), PIC (Spain), ASGC (Taiwan), RAL (UK) and BNL (USA), the Tier-2 facilities worldwide and large non-WLCG resource providers. Major contributors of computing resources are listed in Ref.~\cite{ATL-GEN-PUB-2016-002}.

\bibliographystyle{bibtex/bst/atlasBibStyleWoTitle}
\bibliography{susy,sbottom_Run2,bibtex/bib/ATLAS,bibtex/bib/CMS,bibtex/bib/ConfNotes.bib,bibtex/bib/PubNotes.bib}

\providecommand{\href}[2]{#2}\begingroup\raggedright\begin{thebibliography}{10}

\bibitem{Golfand:1971iw}
{\relax Yu}.~A. Golfand and E.~P. Likhtman, JETP Lett. {\bfseries 13} (1971)
  323--326,
[Pisma Zh. Eksp. Teor. Fiz. 13 (1971) 452].

\bibitem{Volkov:1973ix}
D.~V. Volkov and V.~P. Akulov,
\href{http://dx.doi.org/10.1016/0370-2693(73)90490-5}{Phys. Lett. B {\bfseries
  46} (1973) 109}.

\bibitem{Wess:1974tw}
J.~Wess and B.~Zumino,
\href{http://dx.doi.org/10.1016/0550-3213(74)90355-1}{Nucl. Phys. B {\bfseries
  70} (1974) 39}.

\bibitem{Wess:1974jb}
J.~Wess and B.~Zumino,
\href{http://dx.doi.org/10.1016/0550-3213(74)90112-6}{Nucl. Phys. B {\bfseries
  78} (1974) 1}.

\bibitem{Ferrara:1974pu}
S.~Ferrara and B.~Zumino,
\href{http://dx.doi.org/10.1016/0550-3213(74)90559-8}{Nucl. Phys. B {\bfseries
  79} (1974) 413}.

\bibitem{Salam:1974ig}
A.~Salam and J.~A. Strathdee,
\href{http://dx.doi.org/10.1016/0370-2693(74)90226-3}{Phys. Lett. B {\bfseries
  51} (1974) 353}.

\bibitem{Sakai:1981gr}
N.~Sakai,
\href{http://dx.doi.org/10.1007/BF01573998}{Z. Phys. C {\bfseries 11} (1981)
  153}.

\bibitem{Dimopoulos:1981yj}
S.~Dimopoulos, S.~Raby,  and F.~Wilczek,
\href{http://dx.doi.org/10.1103/PhysRevD.24.1681}{Phys. Rev. D {\bfseries 24}
  (1981) 1681}.

\bibitem{Ibanez:1981yh}
L.~E. Ibanez and G.~G. Ross,
\href{http://dx.doi.org/10.1016/0370-2693(81)91200-4}{Phys. Lett. B {\bfseries
  105} (1981) 439}.

\bibitem{Dimopoulos:1981zb}
S.~Dimopoulos and H.~Georgi,
\href{http://dx.doi.org/10.1016/0550-3213(81)90522-8}{Nucl. Phys. B {\bfseries
  193} (1981) 150}.

\bibitem{Fayet:1976et}
P.~Fayet,
\href{http://dx.doi.org/10.1016/0370-2693(76)90319-1}{Phys. Lett. B {\bfseries
  64} (1976) 159}.

\bibitem{Fayet:1977yc}
P.~Fayet,
\href{http://dx.doi.org/10.1016/0370-2693(77)90852-8}{Phys. Lett. B {\bfseries
  69} (1977) 489}.

\bibitem{Farrar:1978xj}
G.~R. Farrar and P.~Fayet,
\href{http://dx.doi.org/10.1016/0370-2693(78)90858-4}{Phys. Lett. B {\bfseries
  76} (1978) 575}.

\bibitem{Goldberg:1983nd}
H.~Goldberg, \href{http://dx.doi.org/10.1103/PhysRevLett.50.1419}{Phys. Rev.
  Lett. {\bfseries 50} (1983) 1419},
[Erratum: Phys. Rev. Lett. {\bf 103} (2009) 099905].

\bibitem{Ellis:1983ew}
J.~R. Ellis, J.~S. Hagelin, D.~V. Nanopoulos, K.~A. Olive,  and M.~Srednicki,
\href{http://dx.doi.org/10.1016/0550-3213(84)90461-9}{Nucl. Phys. B {\bfseries
  238} (1984) 453}.

\bibitem{Barbieri:1987fn}
R.~Barbieri and G.~F. Giudice,
\href{http://dx.doi.org/10.1016/0550-3213(88)90171-X}{Nucl. Phys. B {\bfseries
  306} (1988) 63}.

\bibitem{deCarlos:1993yy}
B.~de~Carlos and J.~A. Casas,
  \href{http://dx.doi.org/10.1016/0370-2693(93)90940-J}{Phys. Lett. B
  {\bfseries 309} (1993) 320},
\href{http://arxiv.org/abs/hep-ph/9303291}{{\ttfamily arXiv:hep-ph/9303291}}.

\bibitem{SUSY-2014-08}
{ATLAS Collaboration}, \href{http://dx.doi.org/10.1007/JHEP10(2015)134}{JHEP
  {\bfseries 1510} (2015) 134},
  \href{http://arxiv.org/abs/1508.06608}{{\ttfamily arXiv:1508.06608
  [hep-ex]}}.

\bibitem{SUSY-2014-07}
{ATLAS Collaboration},
  \href{http://dx.doi.org/10.1140/epjc/s10052-015-3726-9}{Eur. Phys. J. C
  {\bfseries 75} (2015) 510}, \href{http://arxiv.org/abs/1506.08616}{{\ttfamily
  arXiv:1506.08616 [hep-ex]}}.

\bibitem{CMS-SUS-14-001}
{CMS Collaboration}, \href{http://dx.doi.org/10.1007/JHEP06(2015)116}{JHEP
  {\bfseries 1506} (2015) 116},
  \href{http://arxiv.org/abs/1503.08037}{{\ttfamily arXiv:1503.08037
  [hep-ex]}}.

\bibitem{PERF-2007-01}
{ATLAS Collaboration},
  \href{http://dx.doi.org/10.1088/1748-0221/3/08/S08003}{JINST {\bfseries 3}
  (2008) S08003}.

\bibitem{IBL}
{ATLAS Collaboration},  {ATLAS-TDR-19}, 2010,
  \url{http://cds.cern.ch/record/1291633}, Addendum: ATLAS-TDR-19-ADD-1, 2012,
  \url{http://cds.cern.ch/record/1451888}.

\bibitem{PERF-2011-02}
{ATLAS Collaboration},
  \href{http://dx.doi.org/10.1140/epjc/s10052-011-1849-1}{Eur. Phys. J. C
  {\bfseries 72} (2012) 1849}, \href{http://arxiv.org/abs/1110.1530}{{\ttfamily
  arXiv:1110.1530 [hep-ex]}}.

\bibitem{DAPR-2011-01}
{ATLAS Collaboration},
  \href{http://dx.doi.org/10.1140/epjc/s10052-013-2518-3}{Eur. Phys. J. C
  {\bfseries 73} (2013) 2518}, \href{http://arxiv.org/abs/1302.4393}{{\ttfamily
  arXiv:1302.4393 [hep-ex]}}.

\bibitem{SOFT-2010-01}
{ATLAS Collaboration},
  \href{http://dx.doi.org/10.1140/epjc/s10052-010-1429-9}{Eur. Phys. J. C
  {\bfseries 70} (2010) 823}, \href{http://arxiv.org/abs/1005.4568}{{\ttfamily
  arXiv:1005.4568 [hep-ex]}}.

\bibitem{Agostinelli:2002hh}
{S. Agostinelli et al. (GEANT4)},
\href{http://dx.doi.org/10.1016/S0168-9002(03)01368-8}{Nucl. Instrum. Meth. A
  {\bfseries 506} (2003) 250}.

\bibitem{ATL-PHYS-PUB-2010-013}
{ATLAS Collaboration},  {ATL-PHYS-PUB-2010-013}, 2010,
  \url{http://cds.cern.ch/record/1300517}.

\bibitem{Sjostrand:2007gs}
T.~Sjostrand, S.~Mrenna,  and P.~Z. Skands,
  \href{http://dx.doi.org/10.1016/j.cpc.2008.01.036}{Comput. Phys. Commun.
  {\bfseries 178} (2008) 852},
\href{http://arxiv.org/abs/0710.3820}{{\ttfamily arXiv:0710.3820 [hep-ph]}}.

\bibitem{Martin:2009iq}
A.~D. Martin, W.~J. Stirling, R.~S. Thorne,  and G.~Watt,
  \href{http://dx.doi.org/10.1140/epjc/s10052-009-1072-5}{Eur. Phys. J. C
  {\bfseries 63} (2009) 189},
\href{http://arxiv.org/abs/0901.0002}{{\ttfamily arXiv:0901.0002 [hep-ph]}}.

\bibitem{Martin:2007bv}
A.~Martin, W.~Stirling, R.~Thorne,  and G.~Watt,
  \href{http://dx.doi.org/10.1016/j.physletb.2007.07.040}{Phys. Lett. B
  {\bfseries 652} (2007) 292},
\href{http://arxiv.org/abs/0706.0459}{{\ttfamily arXiv:0706.0459 [hep-ph]}}.

\bibitem{Alwall:2014hca}
J.~Alwall {et~al.}, \href{http://dx.doi.org/10.1007/JHEP07(2014)079}{JHEP
  {\bfseries 1407} (2014) 079},
\href{http://arxiv.org/abs/1405.0301}{{\ttfamily arXiv:1405.0301 [hep-ph]}}.

\bibitem{ATL-PHYS-PUB-2014-021}
{ATLAS Collaboration},  {ATL-PHYS-PUB-2014-021}, 2014,
  \url{http://cdsweb.cern.ch/record/1966419}.

\bibitem{Ball:2012cx}
R.~D. Ball {et~al.},
  \href{http://dx.doi.org/10.1016/j.nuclphysb.2012.10.003}{Nucl. Phys. B
  {\bfseries 867} (2013) 244},
\href{http://arxiv.org/abs/1207.1303}{{\ttfamily arXiv:1207.1303 [hep-ph]}}.

\bibitem{Lonnblad:2012ix}
L.~Lönnblad and S.~Prestel,
  \href{http://dx.doi.org/10.1007/JHEP03(2013)166}{JHEP {\bfseries 1303} (2013)
  166},
\href{http://arxiv.org/abs/1211.7278}{{\ttfamily arXiv:1211.7278 [hep-ph]}}.

\bibitem{Beenakker:1997ut}
W.~Beenakker, M.~Kramer, T.~Plehn, M.~Spira,  and P.~M. Zerwas,
  \href{http://dx.doi.org/10.1016/S0550-3213(98)00014-5}{Nucl. Phys. B
  {\bfseries 515} (1998) 3},
\href{http://arxiv.org/abs/hep-ph/9710451}{{\ttfamily arXiv:hep-ph/9710451}}.

\bibitem{Beenakker:2010nq}
W.~Beenakker {et~al.}, \href{http://dx.doi.org/10.1007/JHEP08(2010)098}{JHEP
  {\bfseries 1008} (2010) 098},
\href{http://arxiv.org/abs/1006.4771}{{\ttfamily arXiv:1006.4771 [hep-ph]}}.

\bibitem{Beenakker:2011fu}
W.~Beenakker {et~al.}, \href{http://dx.doi.org/10.1142/S0217751X11053560}{Int.
  J. Mod. Phys. A {\bfseries 26} (2011) 2637},
\href{http://arxiv.org/abs/1105.1110}{{\ttfamily arXiv:1105.1110 [hep-ph]}}.

\bibitem{Gleisberg:2008ta}
T.~Gleisberg {et~al.},
  \href{http://dx.doi.org/10.1088/1126-6708/2009/02/007}{JHEP {\bfseries 0902}
  (2009) 007}, \href{http://arxiv.org/abs/0811.4622}{{\ttfamily arXiv:0811.4622
  [hep-ph]}}.

\bibitem{Gleisberg:2008fv}
T.~Gleisberg and S.~H{\"o}che,
  \href{http://dx.doi.org/10.1088/1126-6708/2008/12/039}{JHEP {\bfseries 0812}
  (2008) 039}, \href{http://arxiv.org/abs/0808.3674}{{\ttfamily arXiv:0808.3674
  [hep-ph]}}.

\bibitem{Cascioli:2011va}
F.~Cascioli, P.~Maierhofer,  and S.~Pozzorini,
  \href{http://dx.doi.org/10.1103/PhysRevLett.108.111601}{Phys. Rev. Lett.
  {\bfseries 108} (2012) 111601},
\href{http://arxiv.org/abs/1111.5206}{{\ttfamily arXiv:1111.5206 [hep-ph]}}.

\bibitem{Schumann:2007mg}
S.~Schumann and F.~Krauss,
  \href{http://dx.doi.org/10.1088/1126-6708/2008/03/038}{JHEP {\bfseries 0803}
  (2008) 038}, \href{http://arxiv.org/abs/0709.1027}{{\ttfamily arXiv:0709.1027
  [hep-ph]}}.

\bibitem{Hoeche:2012yf}
S.~H{\"o}che, F.~Krauss, M.~Sch{\"o}nherr,  and F.~Siegert,
  \href{http://dx.doi.org/10.1007/JHEP04(2013)027}{JHEP {\bfseries 1304} (2013)
  027},
\href{http://arxiv.org/abs/1207.5030}{{\ttfamily arXiv:1207.5030 [hep-ph]}}.

\bibitem{Lai:2010vv}
H.-L. Lai {et~al.}, \href{http://dx.doi.org/10.1103/PhysRevD.82.074024}{Phys.
  Rev. D {\bfseries 82} (2010) 074024},
\href{http://arxiv.org/abs/1007.2241}{{\ttfamily arXiv:1007.2241 [hep-ph]}}.

\bibitem{Gavin:2010az}
R.~Gavin, Y.~Li, F.~Petriello,  and S.~Quackenbush,
  \href{http://dx.doi.org/10.1016/j.cpc.2011.06.008}{Comput. Phys. Commun.
  {\bfseries 182} (2011) 2388},
\href{http://arxiv.org/abs/1011.3540}{{\ttfamily arXiv:1011.3540 [hep-ph]}}.

\bibitem{PDG:2014}
{K.A. Olive et al. (Particle Data Group)},
  \href{http://dx.doi.org/10.1088/1674-1137/38/9/090001}{Chin. Phys. C
  {\bfseries 38} (2015) 090001}.

\bibitem{Alioli:2010xd}
S.~Alioli, P.~Nason, C.~Oleari,  and E.~Re,
  \href{http://dx.doi.org/10.1007/JHEP06(2010)043}{JHEP {\bfseries 1006} (2010)
  043}, \href{http://arxiv.org/abs/1002.2581}{{\ttfamily arXiv:1002.2581
  [hep-ph]}}.

\bibitem{ATL-PHYS-PUB-2016-004}
{ATLAS Collaboration},  {ATL-PHYS-PUB-2016-004}, 2016,
  \url{http://cdsweb.cern.ch/record/2120417}.

\bibitem{Sjostrand:2006za}
T.~Sjostrand, S.~Mrenna,  and P.~Z. Skands,
  \href{http://dx.doi.org/10.1088/1126-6708/2006/05/026}{JHEP {\bfseries 0605}
  (2006) 026},
\href{http://arxiv.org/abs/hep-ph/0603175}{{\ttfamily arXiv:hep-ph/0603175}}.

\bibitem{Skands:2010ak}
P.~Z. Skands, \href{http://dx.doi.org/10.1103/PhysRevD.82.074018}{Phys. Rev. D
  {\bfseries 82} (2010) 074018},
\href{http://arxiv.org/abs/1005.3457}{{\ttfamily arXiv:1005.3457 [hep-ph]}}.

\bibitem{Czakon:2011xx}
M.~Czakon and A.~Mitov,
  \href{http://dx.doi.org/10.1016/j.cpc.2014.06.021}{Comput. Phys. Commun.
  {\bfseries 185} (2014) 2930},
\href{http://arxiv.org/abs/1112.5675}{{\ttfamily arXiv:1112.5675 [hep-ph]}}.

\bibitem{Kidonakis:2011wy}
N.~Kidonakis, \href{http://dx.doi.org/10.1103/PhysRevD.83.091503}{Phys. Rev. D
  {\bfseries 83} (2011) 091503},
\href{http://arxiv.org/abs/1103.2792}{{\ttfamily arXiv:1103.2792 [hep-ph]}}.

\bibitem{Kidonakis:2010tc}
N.~Kidonakis, \href{http://dx.doi.org/10.1103/PhysRevD.81.054028}{Phys. Rev. D
  {\bfseries 81} (2010) 054028},
\href{http://arxiv.org/abs/1001.5034}{{\ttfamily arXiv:1001.5034 [hep-ph]}}.

\bibitem{Kidonakis:2010ux}
N.~Kidonakis, \href{http://dx.doi.org/10.1103/PhysRevD.82.054018}{Phys. Rev. D
  {\bfseries 82} (2010) 054018},
\href{http://arxiv.org/abs/1005.4451}{{\ttfamily arXiv:1005.4451 [hep-ph]}}.

\bibitem{EvtGen}
{D. J. Lange}, \href{http://dx.doi.org/10.1016/S0168-9002(01)00089-4}{Nucl.
  Instrum. Meth. A {\bfseries 462} (2001) 152}.

\bibitem{ATL-PHYS-PUB-2015-026}
{ATLAS Collaboration},  {ATL-PHYS-PUB-2015-026}, 2015,
  \url{http://cdsweb.cern.ch/record/2037717}.

\bibitem{PERF-2014-07}
{ATLAS Collaboration}, \href{http://arxiv.org/abs/1603.02934}{{\ttfamily
  arXiv:1603.02934 [hep-ex]}}, submitted to Eur. Phys. J. C.

\bibitem{Cacciari:2008gp}
M.~Cacciari, G.~P. Salam,  and G.~Soyez,
  \href{http://dx.doi.org/10.1088/1126-6708/2008/04/063}{JHEP {\bfseries 0804}
  (2008) 063},
\href{http://arxiv.org/abs/0802.1189}{{\ttfamily arXiv:0802.1189 [hep-ph]}}.

\bibitem{ATL-PHYS-PUB-2015-015}
{ATLAS Collaboration},  {ATL-PHYS-PUB-2015-015}, 2015,
  \url{http://cds.cern.ch/record/2037613}.

\bibitem{ATLAS-CONF-2015-037}
{ATLAS Collaboration},  {ATLAS-CONF-2015-037}, 2015,
  \url{http://cdsweb.cern.ch/record/2044941}.

\bibitem{ATLAS-CONF-2015-029}
{ATLAS Collaboration},  {ATLAS-CONF-2015-029}, 2015,
  \url{http://cdsweb.cern.ch/record/2037702}.

\bibitem{ATLAS-CONF-2014-018}
{ATLAS Collaboration},  {ATLAS-CONF-2014-018}, 2014,
  \url{http://cdsweb.cern.ch/record/1700870}.

\bibitem{ATL-PHYS-PUB-2015-039}
{ATLAS Collaboration},  {ATL-PHYS-PUB-2015-039}, 2015,
  \url{http://cdsweb.cern.ch/record/2047871}.

\bibitem{ATLAS-CONF-2014-046}
{ATLAS Collaboration},  {ATLAS-CONF-2014-046}, 2014,
  \url{http://cdsweb.cern.ch/record/1741020}.

\bibitem{PERF-2012-04}
{ATLAS Collaboration},
  \href{http://dx.doi.org/10.1088/1748-0221/11/04/P04008}{JINST {\bfseries 11}
  (2016) P04008},
\href{http://arxiv.org/abs/1512.01094}{{\ttfamily arXiv:1512.01094 [hep-ex]}}.

\bibitem{PERF-2013-03}
{ATLAS Collaboration},
  \href{http://dx.doi.org/10.1140/epjc/s10052-014-2941-0}{Eur. Phys. J. C
  {\bfseries 74} (2014) 2941}, \href{http://arxiv.org/abs/1404.2240}{{\ttfamily
  arXiv:1404.2240 [hep-ex]}}.

\bibitem{PERF-2013-05}
{ATLAS Collaboration},
  \href{http://dx.doi.org/10.1140/epjc/s10052-014-3071-4}{Eur. Phys. J. C
  {\bfseries 74} (2014) 3071}, \href{http://arxiv.org/abs/1407.5063}{{\ttfamily
  arXiv:1407.5063 [hep-ex]}}.

\bibitem{ATL-PHYS-PUB-2015-041}
{ATLAS Collaboration},  {ATL-PHYS-PUB-2015-041}, 2015,
  \url{http://cdsweb.cern.ch/record/2048202}.

\bibitem{PERF-2015-10}
{ATLAS Collaboration}, \href{http://arxiv.org/abs/1603.05598}{{\ttfamily
  arXiv:1603.05598 [hep-ex]}},
submitted to Eur. Phys. J. C.

\bibitem{ATLAS-CONF-2013-082}
{ATLAS Collaboration},  {ATLAS-CONF-2013-082}, 2013,
  \url{http://cdsweb.cern.ch/record/1570993}.

\bibitem{PERF-2011-03}
{ATLAS Collaboration},
  \href{http://dx.doi.org/10.1140/epjc/s10052-013-2304-2}{Eur. Phys. J. C
  {\bfseries 73} (2013) 2304}, \href{http://arxiv.org/abs/1112.6426}{{\ttfamily
  arXiv:1112.6426 [hep-ex]}}.

\bibitem{ATLAS-CONF-2012-123}
{ATLAS Collaboration},  {ATLAS-CONF-2012-123}, 2012,
  \url{http://cdsweb.cern.ch/record/1473426}.

\bibitem{SUSY-2013-05}
{ATLAS Collaboration}, \href{http://dx.doi.org/10.1007/JHEP10(2013)189}{JHEP
  {\bfseries 1310} (2013) 189},
  \href{http://arxiv.org/abs/1308.2631}{{\ttfamily arXiv:1308.2631 [hep-ex]}}.

\bibitem{Tovey:2008ui}
D.~R. Tovey, \href{http://dx.doi.org/10.1088/1126-6708/2008/04/034}{JHEP
  {\bfseries 0804} (2008) 034},
\href{http://arxiv.org/abs/0802.2879}{{\ttfamily arXiv:0802.2879 [hep-ph]}}.

\bibitem{Baak:2014wma}
M.~Baak {et~al.}, \href{http://dx.doi.org/10.1140/epjc/s10052-015-3327-7}{Eur.
  Phys. J. C {\bfseries 75} (2015) 153},
\href{http://arxiv.org/abs/1410.1280}{{\ttfamily arXiv:1410.1280 [hep-ex]}}.

\bibitem{Cowan:2010js}
G.~Cowan, K.~Cranmer, E.~Gross,  and O.~Vitells,
  \href{http://dx.doi.org/10.1140/epjc/s10052-011-1554-0,
  10.1140/epjc/s10052-013-2501-z}{Eur. Phys. J. C {\bfseries 71} (2011) 1554},
  \href{http://arxiv.org/abs/1007.1727}{{\ttfamily arXiv:1007.1727
  [physics.data-an]}},
[Erratum: Eur. Phys. J. C {\bf 73} (2013) 2501].

\bibitem{SUSY-2011-20}
{ATLAS Collaboration},
  \href{http://dx.doi.org/10.1103/PhysRevD.87.012008}{Phys. Rev. D {\bfseries
  87} (2013) 012008}, \href{http://arxiv.org/abs/1208.0949}{{\ttfamily
  arXiv:1208.0949 [hep-ex]}}.

\bibitem{STDM-2011-21}
{ATLAS Collaboration},
  \href{http://dx.doi.org/10.1016/j.physletb.2011.12.046}{Phys. Lett. B
  {\bfseries 707} (2012) 418}, \href{http://arxiv.org/abs/1109.1470}{{\ttfamily
  arXiv:1109.1470 [hep-ex]}}.

\bibitem{Bahr:2008pv}
M.~B{\"a}hr {et~al.},
  \href{http://dx.doi.org/10.1140/epjc/s10052-008-0798-9}{Eur. Phys. J. C
  {\bfseries 58} (2008) 639},
\href{http://arxiv.org/abs/0803.0883}{{\ttfamily arXiv:0803.0883 [hep-ph]}}.

\bibitem{Junk:1999kv}
T.~Junk, \href{http://dx.doi.org/10.1016/S0168-9002(99)00498-2}{Nucl. Instrum.
  Meth. A {\bfseries 434} (1999) 435},
\href{http://arxiv.org/abs/hep-ex/9902006}{{\ttfamily arXiv:hep-ex/9902006}}.

\bibitem{Read:2002hq}
A.~L. Read,
\href{http://dx.doi.org/10.1088/0954-3899/28/10/313}{J. Phys. G {\bfseries 28}
  (2002) 2693}.

\bibitem{SUSY-2013-21}
{ATLAS Collaboration},
  \href{http://dx.doi.org/10.1103/PhysRevD.90.052008}{Phys. Rev. D {\bfseries
  90} (2014) 052008}, \href{http://arxiv.org/abs/1407.0608}{{\ttfamily
  arXiv:1407.0608 [hep-ex]}}.

\bibitem{Aaboud:2016tnv}
{ATLAS Collaboration}, \href{http://arxiv.org/abs/1604.07773}{{\ttfamily
  arXiv:1604.07773 [hep-ex]}},
submitted to Phys. Rev. D.

\bibitem{ATL-GEN-PUB-2016-002}
{ATLAS Collaboration},  {ATL-GEN-PUB-2016-002}, 2016,
  \url{http://cds.cern.ch/record/2202407}.

\end{thebibliography}\endgroup

\newpage 
\begin{flushleft}
{\Large The ATLAS Collaboration}

\bigskip

M.~Aaboud$^\textrm{\scriptsize 135d}$,
G.~Aad$^\textrm{\scriptsize 86}$,
B.~Abbott$^\textrm{\scriptsize 113}$,
J.~Abdallah$^\textrm{\scriptsize 64}$,
O.~Abdinov$^\textrm{\scriptsize 12}$,
B.~Abeloos$^\textrm{\scriptsize 117}$,
R.~Aben$^\textrm{\scriptsize 107}$,
O.S.~AbouZeid$^\textrm{\scriptsize 137}$,
N.L.~Abraham$^\textrm{\scriptsize 149}$,
H.~Abramowicz$^\textrm{\scriptsize 153}$
% ...
\bigskip
\\
% ...
$^{12}$ Institute of Physics, Azerbaijan Academy of Sciences, Baku, Azerbaijan\\
% ...
$^{64}$ University of Iowa, Iowa City IA, United States of America\\
% ...
$^{86}$ CPPM, Aix-Marseille Universit{\'e} and CNRS/IN2P3, Marseille, France\\
% ...
$^{107}$ Nikhef National Institute for Subatomic Physics and University of Amsterdam, Amsterdam, Netherlands\\
% ...
$^{113}$ Homer L. Dodge Department of Physics and Astronomy, University of Oklahoma, Norman OK, United States of America\\
% ...
$^{117}$ LAL, Univ. Paris-Sud, CNRS/IN2P3, Universit{\'e} Paris-Saclay, Orsay, France\\
% ...
$^{135}$ $^{(a)}$ Facult{\'e} des Sciences Ain Chock, R{\'e}seau Universitaire de Physique des Hautes Energies - Universit{\'e} Hassan II, Casablanca; $^{(b)}$ Centre National de l'Energie des Sciences Techniques Nucleaires, Rabat; $^{(c)}$ Facult{\'e} des Sciences Semlalia, Universit{\'e} Cadi Ayyad, LPHEA-Marrakech; $^{(d)}$ Facult{\'e} des Sciences, Universit{\'e} Mohamed Premier and LPTPM, Oujda; $^{(e)}$ Facult{\'e} des sciences, Universit{\'e} Mohammed V, Rabat, Morocco\\
% ...
$^{137}$ Santa Cruz Institute for Particle Physics, University of California Santa Cruz, Santa Cruz CA, United States of America\\
% ...
$^{149}$ Department of Physics and Astronomy, University of Sussex, Brighton, United Kingdom\\
% ...
$^{153}$ Raymond and Beverly Sackler School of Physics and Astronomy, Tel Aviv University, Tel Aviv, Israel\\
% ...
$^{a}$ Also at Department of Physics, King's College London, London, United Kingdom\\
$^{b}$ Also at Institute of Physics, Azerbaijan Academy of Sciences, Baku, Azerbaijan\\
$^{c}$ Also at Novosibirsk State University, Novosibirsk, Russia\\
$^{d}$ Also at TRIUMF, Vancouver BC, Canada\\
$^{e}$ Also at Department of Physics {\&} Astronomy, University of Louisville, Louisville, KY, United States of America\\
% ...
\end{flushleft}


\end{document}